\documentclass[runningheads]{llncs}
\usepackage{graphicx}
\usepackage{tikz}
\usepackage{enumerate}
\usepackage{subfigure}
\usepackage{float}
\newtheorem{branch}{Branching Rule}
\newtheorem{reduction}{Reduction Rule}
\newtheorem{step}{Step}

\begin{document}
\title{Exact algorithms for maximum weighted independent set on sparse graphs} 

\titlerunning{Maximum weighted independent set on sparse graphs} 

\author{Sen Huang\inst{1} \and
Mingyu Xiao\inst{1}\orcidID{0000-0002-1012-2373} \and
Xiaoyu Chen\inst{2}}
\authorrunning{S. Huang et al.}
%
\institute{University of Electronic Science and Technology of China, China,
\email{\{huangsen47,myxiao\}@gmail.com}
\and
Nanjing University, China,
\email{x312035@gmail.com}
}
\maketitle              
\begin{abstract}
The maximum independent set problem is one of the most important problems in graph algorithms and has been extensively studied in the line of research on the worst-case analysis of exact algorithms for NP-hard problems. In the weighted version, each vertex in the graph is associated with a weight and we are going to find an independent set of maximum total vertex weight.
In this paper, we design several reduction rules and a fast exact algorithm for the maximum weighted independent set problem,
and use the measure-and-conquer technique to analyze the running time bound of the algorithm. Our algorithm works on general weighted graphs and it has a good running time bound on sparse graphs. If the graph has an average degree at most 3, our algorithm runs in $O^*(1.1443^n)$ time and polynomial space, improving previous running time bounds for the problem in cubic graphs using polynomial space.


\end{abstract}

\section{Introduction}

The \textsc{Maximum Independent Set} problem on unweighted graphs belongs to the first batch of 21 NP-hard problems proved by Karp~\cite{DBLP:conf/coco/Karp72}.
This problem is so important in graph algorithms that it is often introduced as the first problem in textbooks
and lecture notes of exact algorithms.
In the line of research on the worst-case analysis of exact algorithms for NP-hard problems,
\textsc{Maximum Independent Set}, as one of the most fundamental problems, is used to test the efficiency of new techniques of exact algorithms.

There is a long list of contributions to exact algorithms for \textsc{Maximum Independent Set} in unweighted graphs.
Tarjan and Trojanowski \cite{DBLP:journals/siamcomp/TarjanT77} designed the first nontrivial algorithm in 1977, which runs in $O^*(2^{\frac{n}{3}})$ time and polynomial space.
Later, Jian~\cite{DBLP:journals/tc/Jian86} obtained an $O^*(1.2346^n)$-time algorithm.
Robson~\cite{DBLP:journals/jal/Robson86} gave an $O^*(1.2278^n)$-time polynomial-space algorithm and
an $O^*(1.2109^n)$-time exponential-space algorithm.
By using the measure-and-conquer technique, Fomin et al. \cite{DBLP:journals/jacm/FominGK09} obtained a simple $O^*(1.2210^n)$-time polynomial-space algorithm.
Based on this method, Kneis et al. \cite{DBLP:conf/fsttcs/KneisLR09} and Bourgeois et al. \cite{DBLP:journals/algorithmica/BourgeoisEPR12} improved the running time bound to $O^*(1.2132^n)$ and $O^*(1.2114^n)$, respectively.
Currently, the best algorithm is the $O^*(1.1996^n)$-time polynomial-space algorithm introduced in \cite{DBLP:journals/iandc/XiaoN17}, which even breaks the bound 1.2 in the base of the exponential part of the running time.

For \textsc{Maximum Independent Set} in degree-bounded graphs, there is also a considerable amount of contributions in the literature \cite{DBLP:conf/soda/Beigel99,DBLP:conf/iwpec/BourgeoisEP08,DBLP:journals/algorithmica/ChenKX05,DBLP:journals/tcs/XiaoN13}.
For \textsc{Maximum Independent Set} in degree-3 graphs, let us quote the $O^*(1.1254^n)$-time algorithm by
\cite{DBLP:journals/algorithmica/ChenKX05}, the $O^*(1.1034^n)$-time algorithm by \cite{xiao05},
the $O^*(1.0977^n)$-time algorithm by \cite{DBLP:conf/iwpec/BourgeoisEP08},
the $O^*(1.0892^n)$-time algorithm by \cite{DBLP:journals/jda/Razgon09},
the $O^*(1.0885^n)$-time algorithm by \cite{DBLP:conf/walcom/Xiao10}, the $O^*(1.0854^n)$-time algorithm by \cite{DBLP:journals/algorithmica/BourgeoisEPR12},
the $O^*(1.0836^n)$-time algorithm by \cite{DBLP:journals/tcs/XiaoN13}, and the $O^*(1.0821^n)$-time algorithm by \cite{DBLP:journals/corr/IssacJ13}.
Furthermore, \textsc{Maximum Independent Set} in degree-4 graphs can be solved in $O^*(1.1376^n)$ time \cite{DBLP:journals/jco/XiaoN17}, and
\textsc{Maximum Independent Set} in degree-5 graphs can be solved in $O^*(1.1736^n)$ time \cite{DBLP:journals/dam/XiaoN16}.

In this paper, we will consider the weighted version of \textsc{Maximum Independent Set}, called \textsc{Maximum Weighted Independent Set}, where each vertex in the graph has a nonnegative weight and we are asked to find an independent set with maximum total vertex weight.
Most known results for \textsc{Maximum Weighted Independent Set} were obtained via two counting problems: \textsc{Counting Maximum Weighted Independent Set} and \textsc{Counting Weighted 2SAT}.
Most of these counting algorithms can also list out all independent sets and then we can find a maximum one by increasing only a polynomial factor.
Dahll{\"{o}}f et al.~\cite{DBLP:conf/soda/DahllofJ02} presented an $O^*(1.3247^n)$-time algorithm for \textsc{Counting Maximum Weighted Independent Set}. Later, the running time bound was improved to $O^*(1.2431^n)$ by Fomin et al.~\cite{DBLP:conf/isaac/FominGS06}.
\textsc{Counting Maximum Weighted Independent Set} can also be reduced to \textsc{Counting Weighted 2SAT},
preserving the exponential part of the running time. In the reduction,
we  construct a clause $\overline{u}\lor \overline{v}$ for each edge $uv$ in the graph (See~\cite{DBLP:journals/tcs/DahllofJW05} for more details).
For \textsc{Counting Weighted 2SAT},
the running time bound was improved from $O^*(1.2561^n)$~\cite{DBLP:journals/tcs/DahllofJW05} to $O^*(1.2461^n)$~\cite{DBLP:conf/aaim/FurerK07} and then to
$O^*(1.2377^n)$~\cite{DBLP:conf/iwpec/Wahlstrom08}.
 Wahlstr{\"{o}}m~\cite{DBLP:conf/iwpec/Wahlstrom08} also showed that the running time bound could be further improved to $O^*(1.1499^n)$ and $O^*(1.2117^n)$ if the maximum degree of the variables or the vertices in the graph is bounded by 3 and 4, respectively. Most of the above algorithms use only polynomial space.
 If exponential space is allowed, dynamic programming algorithms based on tree decompositions can achieve a better running time bound.
 On graphs of treewidth at most $t$, \textsc{Maximum Weighted Independent Set} can be solved in $O^*(2^t)$ time and space by a standard dynamic programming algorithm.
 It is known that the treewidth of graphs with maximum degree 3 and 4 is roughly bounded by $n/6$ and $n/3$, respectively~\cite{DBLP:journals/algorithmica/FominGSS09}.
 Thus, \textsc{Maximum Weighted Independent Set} in graphs of maximum degree 3 (resp., 4) can be solved in $O^*(1.1225^n)$ time (resp., $O^*(1.2600^n)$ time) and exponential space.
 We also note that the due problem \textsc{Minimum Weighted Vertex Cover} has been extensively studied in parameterized complexity.
By taking the weight value $W$ of the vertex cover as the parameter, some parameterized algorithms have been proposed in \cite{DBLP:journals/jal/NiedermeierR03} and \cite{DBLP:conf/isaac/FominGS06}.
By taking the size $s$ of the minimum weighted vertex cover as the parameter, there are also some known parameterized algorithms \cite{DBLP:journals/jcss/ShachnaiZ17}.

\textsc{Maximum Weighted Independent Set} is an important problem with many applications in various real-world problems. For example, the dynamic map labeling problem \cite{DBLP:journals/tvcg/BeenDY06,DBLP:journals/tcs/LiaoLP16} can be naturally encoded as \textsc{Maximum Weighted Independent Set}.
Some experimental algorithms, such as the algorithms in~\cite{DBLP:conf/alenex/Lamm0SWZ19,DBLP:conf/www/XiaoHZD21} have been developed to solve instances from real world and known benchmarks. These algorithms run fast even on large scale sparse instances but lack running time analysis.
On the other hand, the fast algorithms for \textsc{Counting Maximum Weighted Independent Set} and \textsc{Counting Weighted 2SAT}
and the DP algorithm based tree decompositions do not rely on the structural properties of \textsc{Maximum Weighted Independent Set}.

 In this paper, we will focus on exact algorithms specifying
 for \textsc{Maximum Weighted Independent Set}. We develop structural properties and design reduction rules for the problem, and then design a fast exact algorithm based on them.
 By using the measure-and-conquer technique, we can prove that the algorithm runs in $O^*(1.1443^{(0.624x-0.872)n})$ time and polynomial space, where $x$ is the average degree of the graph. For some sparse graphs, our result beats the known bounds. The running time bound of our algorithm in graphs with the average degree at most three is $O^*(1.1443^n)$, which improves the previously known bound of $O^*(1.1499^n)$ using polynomial space \cite{DBLP:conf/iwpec/Wahlstrom08}. When the average degree is slightly greater than three, we may still get a good running time bound.
 For graphs with the average degree at most 3.68, the running time of our algorithm is strictly better than the running time bound $O^*(1.2117^n)$ for \textsc{Maximum Weighted Independent Set} in degree-4 graphs \cite{DBLP:conf/iwpec/Wahlstrom08}.

\section{Preliminaries}

Let $G=(V,E,w)$ denote an undirected vertex-weighted graph with $|V|=n$ vertices and $|E|=m$ edges, where
each vertex $v\in V$ is associated with a positive weight $w_G(v)$, where the subscript $G$ may be omitted if it is clear from the context.
Although our graphs are undirected, we may use an arc to denote the relation of the weights of the two endpoints of an edge.
An \emph{arc} $\overrightarrow{uv}$ from vertex $u$ to vertex $v$ means that there is an edge between $u$ and $v$ and it holds that  $w(u)\geq w(v)$.

For a vertex subset $V'\subseteq V$, we let $w(V')=\sum_{v\in V'}w(v)$.
For a vertex subset $V'\subseteq V$, we let $N_G(V')$ denote the \emph{open neighborhood} $V'$, i.e., $N_G(V')=\{v| \mbox{$v$ is adjacent to some}$ $\mbox{vertex in $V'$} ~\&~ v\not\in V'\}$.
We also let $d_G(V')=|N_G(V')|$ and $N_G[V']=N_G(V')\cup V'$, where $N_G[V']$ is the \emph{closed neighborhood} of $V'$.
When the graph $G$ is clear from the context, we may omit the subscript and simply write $N_G(V')$, $d_G(V')$ and $N_G[V']$ as $N(V')$, $d(V')$ and $N[V']$, respectively. When $V'=\{v\}$ is a singleton, we may simply write it as $v$.
We also use $G[V']$ to denote the subgraph of $G$ induced by $V'$ and use $G-V'$ to denote $G[V\setminus V']$. For a graph $G'$, we use $\mathcal{C}(G')$ to denote the set of connected components of $G'$.
A \emph{chain} is an induced path such that the degree of each vertex except the two endpoints of the path is exactly 2.
One vertex is a \emph{chain-neighbor} of another vertex if they are connected by a chain.

A path (or cycle) is said to be a $k$-path (or $k$-cycle) if there are $k$ edges.
A set $S$ of vertices in graph $G$ is called an \emph{independent set} if for any pair of vertices in $S$ there is no edge between them.
For a vertex-weighted graph, a \emph{maximum weighted independent set} is an independent set $S$ such that $w(S)$ is maximized among all independent sets in the graph.
We use $S(G)$ to denote a  maximum weighted independent set in graph $G$ and $\alpha(G)$ to denote the total vertex weight of $S(G)$.
The \textsc{Maximum Weighted Independent Set} problem is defined below.

\noindent\rule{\linewidth}{0.2mm}
\textsc{Maximum Weighted Independent Set (MWIS)}\\
\textbf{Input}: An undirected vertex-weighted graph $G=(V,E,w)$.\\
\textbf{Output}: the weight of a maximum weighted independent set in $G$., i.e., $\alpha(G)$. \\
\rule{\linewidth}{0.2mm}

\subsection{Branch-and-search and Measure-and-conquer}
\textbf{Branch-and-search paradigm.}  Our branch-and-search algorithm contains several reduction rules and branching rules. Each reduction rule will reduce the instance without exponentially increasing the running time. We will first apply reduction rules to reduce this instance and then apply branching rules to search for a solution when the instance can not be further reduced.

The exponential part of the running time depends on the size of the ``search tree''
in the algorithm, which is generated by the branching operations. To evaluate the size of the search tree, we should use a measure. The measure can be the number of vertices or edges
of the graph, the size of the solution, and so on.  Usually, when the parameter becomes zero or less than zero, the instance can be solved in polynomial time directly.
Let parameter $p$ be the measure adopted in the algorithm.
We use $T(p)$ to denote the maximum number of leaves in the search tree generated in the algorithm for any instance with the measure being at most $p$.
Assume that at a branching operation, the algorithm branches on the current instance into $l$ branches.
If in the $i$-th branch the measure decreases by at least $a_i$, i.e., the $i$-th substance has the parameter at most $p_i=p-a_i$, then we obtain a recurrence relation
$$T(p)\leq T(p-a_1)+T(p-a_2)+\dots + T(p-a_l).$$
The recurrence relation can be represented by a \emph{branching vector} $[a_1,a_2,\dots,a_l]$.
The largest root of the function $f(x) = 1-\sum_{i=1}^lx^{-a_i}$ is called the \emph{branching factor} of the recurrence.
Let $\gamma$ be the maximum branching factor among all branching factors in the algorithm.
The size of the search tree that represents the branching process of the algorithm applied to an instance with parameter $p$ is given by $O^*(\gamma^p)$. More details about the analysis and how to solve recurrences can be found in the monograph~\cite{DBLP:series/txtcs/FominK10}.

For two branching vectors $\mathbf{a}=[a_1,a_2,\dots,a_l]$ and $\mathbf{b}=[b_1,b_2,\dots,b_l]$, if $a_i\geq b_i$ holds for all $i=1,2\dots,l$, then the branching factor of $\mathbf{a}$ is not greater than this of $\mathbf{b}$.
For this case, we say $\mathbf{b}$ \emph{dominates} $\mathbf{a}$.
This property will be used in many places to simplify some arguments in the paper.

\medskip
\noindent\textbf{Measure-and-conquer technique.}
The Measure-and-conquer technique, introduced in~\cite{DBLP:journals/jacm/FominGK09}, is a powerful tool to analyze branch-and-search algorithms.
The main idea of the measure-and-conquer technique is to use a non-traditional measure to evaluate the size of the search tree generated by the branch-and-search algorithm.
In this paper, we will use the measure-and-conquer technique to analyze our algorithm.
Our measure $p$ is a combination of several parameters defined below. This measure may catch more structural properties of the problem and then we can analyze the running time by using amortization.
Let $n_i$ denote the number of vertices of degree $i$ in the graph. We associate a cost $\delta_i\geq 0$ for
each degree-$i$ vertex in the graph. Our measure is set as follows:
\begin{eqnarray}\label{e_themeasure}
p:=\sum_{i=0}^nn_i\delta_i.
\end{eqnarray}

The cost $\delta_i$ in this paper is given by
\begin{eqnarray} \label{e_costsetting}
\delta_i {\rm{ = }}\left\{ {\begin{array}{*{20}{l}}
0&{{\rm{if}} ~i \leq 1}\\
{0.376}&{{\rm{if}} ~i=2}\\
1&{{\rm{if}} ~i=3}\\
{1+0.624(i-3)}&{\rm{if}} ~i \geq 4.
\end{array}} \right.
\end{eqnarray}

We also define
$$\delta_i^{<-k>}:=\delta_i-\delta_{i-k},$$
for each integer $k\geq 0$.
In our analysis, we may use the following inequalities and equalities to simplify some arguments:
\begin{eqnarray}\delta_i^{<-1>}=\delta_3^{<-1>} \mbox{~~~~for each~~} i\geq 4;\end{eqnarray}
\begin{eqnarray}\delta_3\geq 2.5\delta_2;\end{eqnarray}
\begin{eqnarray}3\delta_2\geq \delta_3.\end{eqnarray}

With the above setting, we know that when $p\leq0$, the instance contains only degree-0 and degree-1 vertices and can be solved
directly. We will design an algorithm with running time bound $O^*(c^p)$ for some constant $c$.
If the initial graph has degree at most 3, then we have that $p\leq n$ and then the running time bound of the algorithm is $O^*(c^n)$.
In general, if we have $p\leq f(n)$ for some function $f$ on $n$, then we can get a running time bound of $O^*(c^{f(n)})$.
We have the following lemma for the relation between $p$ and $n$.

\begin{lemma}\label{themeasure}
    For a graph of $n$ vertices, if the average degree of the graph is at most $x$, then the measure $p$ of the graph is at most $(0.624x-0.872)n$.
    \end{lemma}
    \begin{proof}
    According to the definition of the measure, we have that
    $$p=\sum_{i=1}^nn_i\delta_i=\sum_{i=2}^nn_i\delta_i$$
    $$= \sum_{i=2}^nn_i\delta_2 +\sum_{i=2}^nn_i(i-2)\delta_3^{<-1>}.$$
    Since the average degree of the graph is at most $x$, we have that
    $$ \sum_{i=0}^nn_ii\leq x\sum_{i=0}^nn_i.$$
    Thus,
    $$p\leq \sum_{i=2}^nn_i\delta_2 + \sum_{i=0}^nn_i(x-2)\delta_3^{<-1>}\leq n(\delta_2+(x-2)\delta_3^{<-1>})$$
    $$=n+(x-3)\delta_3^{<-1>}n= (0.624x-0.872)n.$$
\end{proof}

\section{Reduction Rules}
We first introduce some reduction rules, which can be applied to reduce the instance directly by eliminating some local structures of the graph. Reduction rules for the unweighted case have been extensively studied.
However, most of they do not work in weighted graphs. In weighted graphs, we may not be able to reduce all degree-2 vertices, which
is an easy case in unweighted graphs. There are also two papers~\cite{DBLP:conf/alenex/Lamm0SWZ19,DBLP:conf/www/XiaoHZD21} systematically study reduction rules for the weighted case. Here we try to contribute more reduction rules based on degree-2 vertices, small vertex-cuts, and some other special local structures.

Some reduction rules may include a set $S$ of vertices in the solution set directly.
We use $M_c$ to store the weight of the vertices that have been included in the solution set.
When a set $S$ of vertices is included in the solution set, we will remove $N[S]$ from the graph and update $M_c$ by adding $w(S)$.

\subsection{General Reductions for Some Special Structures}
We use several reduction rules based on unconfined vertices, twins, vertices with a clique neighborhood, and heavy vertices.
Some of these reduction rules were introduced in \cite{DBLP:conf/alenex/Lamm0SWZ19} and \cite{DBLP:conf/www/XiaoHZD21}.

\medskip
\noindent\textbf{Unconfined Vertices.} A vertex $v$ in $G$ is called \emph{removable} if $\alpha(G)=\alpha(G-v)$, i.e., there is a maximum weighted independent set in $G$ that does not contain $v$.
We can say that a vertex $v$ is removable if a contradiction is obtained from the assumption that every maximum weighted independent set in $G$ contains $v$.
A sufficient condition for a vertex to be removable in unweighted graphs has been studied in \cite{DBLP:journals/tcs/XiaoN13}.
We extend this concept to weighted graphs.

\begin{figure}
    \centering
    \begin{tikzpicture}
        [scale = 1, line width = 0.5pt,solid/.style = {circle, draw, fill = black, minimum size = 0.3cm},empty/.style = {circle, draw, fill = white, minimum size = 0.5cm}]
        \draw[dashed]  (0, 0) ellipse(1 and 0.5);

        \node[empty] (A) at (0.5,0) {};
        \node[empty] (B) at (-0.5,0) {};

        \node[] (C) at (-1.5,0) {$S$};

        \node[empty,label=center:$u'$] (E) at (0.5,-1) {};
        \node[empty] (F) at (-0.5,-1) {$u$};
        \node[empty] (G) at (1.5,-1) {};
        \node[empty] (H) at (-1,-2) {$v$};

        \draw (A)--(F);
        \draw (A)--(E);
        \draw (A)--(G);

        \draw (B)--(F);
        \draw (E)--(G);
        \draw (E)--(0.15,-1.6);
        \draw (E)--(0.9,-1.6);
        \draw (G)--(1.9,-1.6);

        \draw (F)--(H);
        \draw (H)--(-1.4,-2.6);
        \draw (H)--(-0.6,-2.6);
    \end{tikzpicture}
    \caption{An independent $S$ of $G$, where vertices $u$ and $u'$ are children of $S$, vertex $u$ is an extending child and $v$ is a satellite of $S$}
    \label{satellite}
\end{figure}
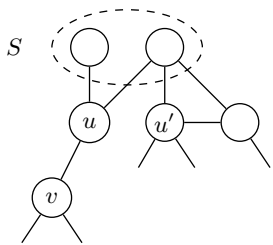

For an independent set $S$ of $G$, a vertex $u\in N(S)$ is called a \emph{child} of $S$ if $w(u)\geq w(S\cap N(u))$.
A child $u$ is called an \emph{extending child} if it holds that $|N(u)\setminus N[S]|=1$, and the only vertex $v\in N(u)\setminus N[S]$
is called a \emph{satellite} of $S$.
See Fig.~\ref{satellite} for an illustration.

\begin{lemma}\label{lem_1}
    Let $S$ be an independent set that is contained in any maximum weighted independent set in $G$.
    Then every maximum weighted independent set contains at least one vertex in $N(u)\setminus N[S]$ for each child $u$ of $S$.
\end{lemma}
\begin{proof}
    Assume to the contrary that there is a maximum weighted independent set $S_G$ in $G$ such that $S_G\cap (N(u)\setminus N[S])=\emptyset$ for some child $u$ of $S$.
    We have that $S\subseteq S_G$ by the assumption on the independent set $S$.
    Thus, we could replace $S\cap N(u)$ with $u$ to obtain another independent set $S'_{G}=(S_G\setminus N(u))\cup\{u\}$ in $G$.
    Furthermore, since $u$ is a child of $S$, we have that $w(u)\geq w(S\cap N(u))$.
    Thus, $S'_{G}$ is also a maximum weighted independent set in $G$.
    However $S_G'$ does not contain $S$, contradicting that $S$ is contained in any maximum weighted independent set in $G$.
\end{proof}

Lemma~\ref{lem_1} provides a sufficient condition for a vertex set contained in any maximum weighted independent set.
Next, we introduce a method based on Lemma~\ref{lem_1} to find some possible removable vertices.

Let $v$ be an arbitrary vertex in the graph. After starting with $S:=\{v\}$, we repeat (1) until (2) or (3) holds:

\begin{enumerate}[(1)]
    \item If $S$ has some extending child in $N(S)$, then let $S'$ be the set of satellites.
    Update $S$ by letting $S:=S\cup S'$.
    \item If $S$ is not an independent set or there is a child $u$ such that $N(u)\setminus N[S]=\emptyset$, then halt and conclude that $v$ is \emph{unconfined}.
    \item If $|N(u)\setminus N[S]|\geq 2$ for all children $u\in N(S)$, then halt and return $S_v=S$.
\end{enumerate}

Obviously, the procedure can be executed in polynomial time for any starting set $S$ of a vertex.
If the procedure halts in (2), we say vertex $v$ \emph{unconfined}.
If the procedure halts in (3), then we say that the set $S_v$ returned in (3) \emph{confines} vertex $v$ and vertex $v$ is also called \emph{confined}. Note that the set $S_v$ confining $v$ is uniquely determined by the procedure with starting set $S:=\{v\}$. It is easy to observe the following lemma.

\begin{lemma}
Any unconfined vertex is removable.
\end{lemma}
\begin{proof}
Assume that vertex $v$ is contained in all maximum weighted independent sets.
By Lemma~\ref{lem_1}, we know that after each execution of (1) of the above procedure, the set $S$ should also be
contained in all maximum weighted independent sets.
However, when the procedure halts in (2), the final set $S$ could not be contained in any maximum weighted independent set
by Lemma~\ref{lem_1}. So $v$ is removable.
\end{proof}

\begin{reduction}[R1]
    \label{remove_ucf}
    If a vertex $v$ is unconfined, remove $v$ from $G$.
\end{reduction}

\begin{figure}
    \centering
    \subfigure{
        \begin{tikzpicture}
            [scale = 0.7, line width = 0.5pt,solid/.style = {circle, draw, fill = black, minimum size = 0.3cm},empty/.style = {circle, draw, fill = white, minimum size = 0.5cm}]
            \draw (0, 0) ellipse (2 and 1);
            \node[empty] (A) at (-1.5,2) {$u$};
            \node[empty] (B) at (1.5,2) {$v$};
            \draw (A) -- (-1,0);
            \draw (A) -- (1,0);
            \draw (B) -- (-1,0);
            \draw (B) -- (1,0);
            \draw[->] (A) -- (B);
            \draw (B) -- (0,0);
        \end{tikzpicture}
    }
    \qquad\qquad\qquad
    \subfigure{
        \begin{tikzpicture}
            [scale = 0.7, line width = 0.5pt,solid/.style = {circle, draw, fill = black, minimum size = 0.3cm},empty/.style = {circle, draw, fill = white, minimum size = 0.5cm}]
            \draw (0, 0) ellipse (2 and 1);
            \node[empty] (A) at (-1.5,2) {$u$};
            \draw (A) -- (-1,0);
            \draw (A) -- (1,0);
        \end{tikzpicture}
    }

    \caption{(a) the graph $G$, where vertex $v$ is dominated by vertex $u$, and $v$ is removable; (b) the graph $G'$ that is obtained from $G$ by deleting the removable vertex $v$}
    \label{domination}
\end{figure}
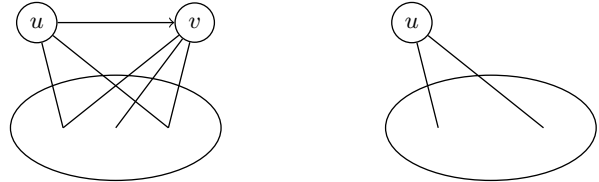

We here observed some structures that are involved in unconfined vertices.
We say that a vertex $v$ \emph{dominated} by a neighbor $u$ of it if $v$ is adjacent to all neighbors of $u$, i.e., $N[u]\subseteq N[v]$.
Clearly, any dominated vertex $v$ with $w(v)\leq w(u)$ is unconfined, since $S=\{v\}$ has a child $u$ with $N(u)\setminus N[S]=\emptyset$. See Fig.~\ref{domination} for an illustration of this case.
For a degree-1 vertex $u$ with the unique neighbor $v$, if $w(v)\leq w(u)$, then vertex $v$ is unconfined.
For a degree-2 vertex $u$ with two adjacent neighbors, if one neighbor, say $v$, holds that $w(v)\leq w(u)$, then vertex $v$ is unconfined. R\ref{remove_ucf} can deal with some degree-1 and degree-2 vertices, but not all of them.

\medskip
\noindent\textbf{Twins.}
    A set $A=\{u,v\}$ of two non-adjacent vertices is called a \emph{twin} if they have the same neighbor set, i.e., $N(u)=N(v)$. Reductions based on twins are used not only for independent sets~\cite{DBLP:journals/tcs/AkibaI16,DBLP:conf/alenex/Lamm0SWZ19} but also for feedback sets and other problems~\cite{DBLP:journals/dam/McConnellM05}. Clearly, a vertex in a twin is in a maximum weighted independent set if and only if the other vertex in the twin is also in the same maximum weighted independent set. So we can treat the two vertices in a twin as a single vertex.

\begin{reduction}[R2] \cite{DBLP:conf/alenex/Lamm0SWZ19}
    \label{fold_twin}
    If there is a twin $A=\{u,v\}$, delete $v$ and update the weight of  $u$ by letting $w(u):=w(u)+w(v)$.
\end{reduction}

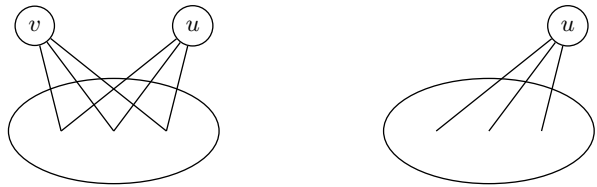
\begin{figure}
    \centering
    \subfigure{
        \centering
        \begin{tikzpicture}
            [scale = 0.7, line width = 0.5pt,solid/.style = {circle, draw, fill = black, minimum size = 0.3cm},empty/.style = {circle, draw, fill = white, minimum size = 0.5cm}]
            \draw (0, 0) ellipse (2 and 1);
            \node[empty] (A) at (-1.5,2) {$v$};
            \node[empty] (B) at (1.5,2) {$u$};
            \draw (A) -- (-1,0);
            \draw (A) -- (1,0);
            \draw (A) -- (0,0);
            \draw (B) -- (-1,0);
            \draw (B) -- (1,0);
            \draw (B) -- (0,0);
        \end{tikzpicture}
    }
    \qquad\qquad\qquad
    \subfigure{
        \centering
        \begin{tikzpicture}
            [scale = 0.7, line width = 0.5pt,solid/.style = {circle, draw, fill = black, minimum size = 0.3cm},empty/.style = {circle, draw, fill = white, minimum size = 0.5cm}]
            \draw (0, 0) ellipse (2 and 1);
            \node[empty] (B) at (1.5,2) {$u$};
            \draw (B) -- (-1,0);
            \draw (B) -- (1,0);
            \draw (B) -- (0,0);
        \end{tikzpicture}
    }
    \caption{(a) the graph $G$ that has a twin $\{u,v\}$; (b) the graph $G'$ after deleting $v$ from $G$ and updating the weight of  $u$ by letting $w(u)=w(u)+w(v)$}
    \label{twins}
\end{figure}

See Fig.~\ref{twins} for an illustration of R\ref{fold_twin}.

\medskip
\noindent\textbf{Clique Neighborhood}.
    A vertex $v$ has a \emph{clique neighborhood} if the graph $G[N(v)]$ induced by the open neighbor set of $v$ is a clique, which was introduced as isolated vertices in~\cite{DBLP:conf/alenex/Lamm0SWZ19}.

\begin{reduction}[R3] \cite{DBLP:conf/alenex/Lamm0SWZ19}
    \label{neighbor_clique}
    If there is  a vertex $v$ having a clique neighborhood and $w(v)<w(u)$ holds for all $u\in N(v)$, then remove $v$ from the graph, update the weight $w(u):=w(u)-w(v)$ for all $u\in N_G(v)$, and add $w(v)$ to $M_c$.
\end{reduction}

\begin{figure}
    \centering
    \subfigure{
        \centering
        \begin{tikzpicture}
            [scale = 0.7, line width = 0.5pt,solid/.style = {circle, draw, fill = black, minimum size = 0.3cm},empty/.style = {circle, draw, fill = white, minimum size = 0.5cm}]
            \draw (0, 0) ellipse (2 and 1);
            \node[empty,label=center:$u_1$] (A) at (-1.5,2) {};
            \node[empty,label=center:$u_2$] (B) at (0,1.5) {};
            \node[empty,label=center:$u_3$] (C) at (1.5,2) {};
            \node[empty] (D) at (0,3) {$v$};
            \draw (A) -- (-1,0);
            \draw (C) -- (1,0);
            \draw (B) -- (0,0);
            \draw (A) -- (B);
            \draw (C) -- (B);
            \draw (A) -- (C);
            \draw[->] (A) -- (D);
            \draw[->] (B) -- (D);
            \draw[->] (C) -- (D);
        \end{tikzpicture}
    }
    \qquad\qquad\qquad
    \subfigure{
        \centering
        \begin{tikzpicture}
            [scale = 0.7, line width = 0.5pt,solid/.style = {circle, draw, fill = black, minimum size = 0.3cm},empty/.style = {circle, draw, fill = white, minimum size = 0.5cm}]
            \draw (0, 0) ellipse (2 and 1);
            \node[empty,label=center:$u_1$] (A) at (-1.5,2) {};
            \node[empty,label=center:$u_2$] (B) at (0,1.5) {};
            \node[empty,label=center:$u_3$] (C) at (1.5,2) {};
            \draw (A) -- (-1,0);
            \draw (C) -- (1,0);
            \draw (B) -- (0,0);
            \draw (A) -- (B);
            \draw (C) -- (B);
            \draw (A) -- (C);
        \end{tikzpicture}

    }
    \caption{(a) the graph $G$, where vertex $v$ has a clique neighborhood and the weight of $v$ is less than the weight of any neighbor of it; (b) the graph $G'$ after deleting $v$ from $G$ and updating the weight $w(u):=w(u)-w(v)$ for all $u\in N_G(v)$}
    \label{cliqueneighbor}
\end{figure}
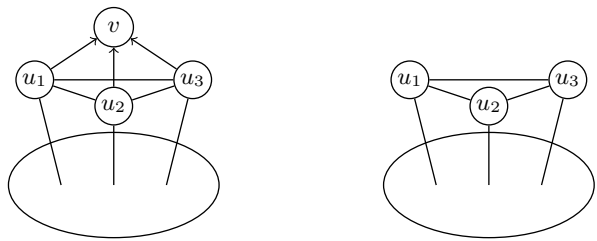

    An illustration of which is shown in Fig.~\ref{cliqueneighbor}.

\medskip
\noindent\textbf{Heavy Vertices}. A vertex $v$ is called a \emph{heavy vertex} if its weight is not less the weight of the maximum weighted independent set in subgraph induced by the open neighborhood of it, i.e.,
$$
w(v)\geq \alpha(G[N(v)]).
$$

We can see that for each heavy vertex, there is a maximum weighted independent set contain it. Note that if there is a maximum weighted independent set $S$ does not contain $v$, then we can replace $N(v)\cap S$ with $v$ in $S$ to get
another maximum weighted independent set, where $w(N(v)\cap S)\leq w(v)$ since $v$ is a heavy vertex.
 Dealing heavy vertices is a simple but very efficient method to reduce the graph that have been used in some experimental algorithms~\cite{DBLP:conf/alenex/Lamm0SWZ19,DBLP:conf/www/XiaoHZD21}.
Whether a vertex is a heavy vertex can be checked in constant time if the degree of the vertex is bounded by a constant.
In this paper, we will only check heavy vertices of degree bounded by 5. Note that degree-0 vertices are heavy vertices and we can reduce degree-0 vertices in this step.

\begin{reduction}[R4]
    \label{heavy_vertices}
If there is a heavy vertex $v$ of degree at most 5, then delete $N[v]$ from the graph and add $w(v)$ to $M_c$.
\end{reduction}

\subsection{Reductions Based on Degree-2 Vertices}

For unweighted graphs, we have good reduction rules to deal with all degree-2 vertices (see the reduction rule in~\cite{DBLP:journals/jal/ChenKJ01}).
However, for weighted graphs, it becomes much more complicated. Although we have several reduction rules, we can not deal with all degree-2 vertices.

     Our first rule is generalized from the concept of folding degree-2 vertices in unweighted graphs introduced in \cite{DBLP:journals/jal/ChenKJ01}, which has been also used in some experimental algorithms. A proof of the correctness can be found in~\cite{DBLP:conf/alenex/Lamm0SWZ19,DBLP:conf/www/XiaoHZD21}.

\begin{reduction}[R5]
    \label{fold_d_2_2}
    If there is a degree-2 vertex $v$ with two neighbors $\{u_1,u_2\}$ such that $w(u_1)+w(u_2)>w(v)\geq max\{w(u_1),w(u_2)\}$,
    then delete $\{v, u_1,u_2\}$ from the graph $G$, introduce a new vertex $v'$ adjacent to $N_G(\{v,u_1,u_2\})$ with weight $w(v'):=w(u_1)+w(u_2)-w(v)$, and add $w(v)$ to $M_c$.
\end{reduction}

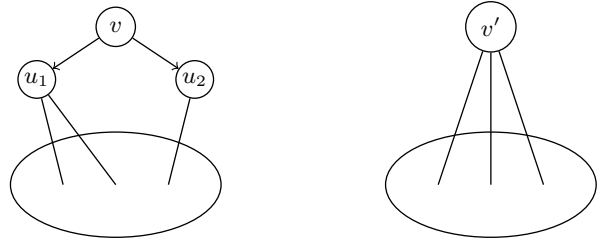
\begin{figure}
    \centering
    \subfigure{
        \centering
        \begin{tikzpicture}
            [scale = 0.7, line width = 0.5pt,solid/.style = {circle, draw, fill = black, minimum size = 0.3cm},empty/.style = {circle, draw, fill = white, minimum size = 0.5cm}]
            \draw (0, 0) ellipse (2 and 1);
            \node[empty,label=center:$u_1$] (A) at (-1.5,2) {};
            \node[empty,label=center:$u_2$] (C) at (1.5,2) {};
            \node[empty] (D) at (0,3) {$v$};
            \draw (A) -- (-1,0);
            \draw (A) -- (0,0);
            \draw (C) -- (1,0);
            \draw[->] (D) -- (A);
            \draw[->] (D) -- (C);
        \end{tikzpicture}

    }
    \qquad\qquad\qquad
    \subfigure{
        \centering
        \begin{tikzpicture}
            [scale = 0.7, line width = 0.5pt,solid/.style = {circle, draw, fill = black, minimum size = 0.3cm},empty/.style = {circle, draw, fill = white, minimum size = 0.5cm}]
            \draw (0, 0) ellipse (2 and 1);
            \node[empty] (D) at (0,3) {$v'$};
            \draw (D) -- (-1,0);
            \draw (D) -- (0,0);
            \draw (D) -- (1,0);
        \end{tikzpicture}
    }

    \caption{(a) the graph $G$ having a degree-2 vertex $v$ with two neighbors $u_1$ and $u_2$; (b) the graph $G'$ after deleting $\{v, u_1,u_2\}$ and introducing the new vertex $v'$}
    \label{rr2-2}
\end{figure}

Fig.~\ref{rr2-2} gives an illustration of R\ref{fold_d_2_2}.

The following two reductions are special cases of alternative sets introduced in~\cite{DBLP:conf/www/XiaoHZD21}. We also use them to reduce some degree-2 vertices in the graph.

\begin{reduction}[R6]
    \label{3-path}
    If there is a path $v_1v_2v_3v_4$ such that $d_G(v_2)=d_G(v_3)=2$ and $w(v_1)\geq w(v_2)\geq w(v_3)\geq w(v_4)$,
    then remove $v_2$ and $v_3$ from the graph, add an edge $v_1v_4$ if it does not exist, update the weight of $v_1$ by letting
    $w(v_1):=w(v_1)+w(v_3)-w(v_2)$, and add $w(v_2)$ to $M_c$.
\end{reduction}

Fig.~\ref{rr2-3} gives an illustration of R\ref{3-path}. A proof of the correctness of this reduction rule is given below.

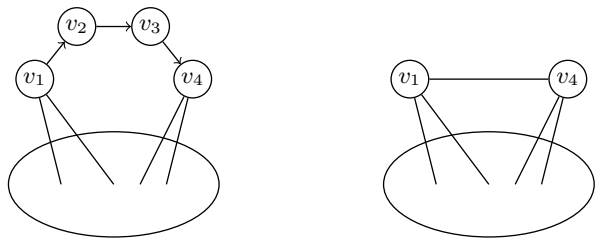
\begin{figure}
    \centering
    \subfigure{
        \centering
        \begin{tikzpicture}
            [scale = 0.7, line width = 0.5pt,solid/.style = {circle, draw, fill = black, minimum size = 0.3cm},empty/.style = {circle, draw, fill = white, minimum size = 0.5cm}]
            \draw (0, 0) ellipse (2 and 1);
            \node[empty,label=center:$v_1$] (A) at (-1.5,2) {};
            \node[empty,label=center:$v_2$] (B) at (-0.7,3) {};
            \node[empty,label=center:$v_3$] (C) at (0.7,3) {};
            \node[empty,label=center:$v_4$] (D) at (1.5,2) {};
            \draw (A) -- (-1,0);
            \draw (D) -- (1,0);
            \draw (D) -- (0.5,0);
            \draw (A) -- (0,0);
            \draw[->] (A) -- (B);
            \draw[->] (B) -- (C);
            \draw[->] (C) -- (D);
        \end{tikzpicture}
    }
    \qquad\qquad\qquad
    \subfigure{
        \centering
        \begin{tikzpicture}
            [scale = 0.7, line width = 0.5pt,solid/.style = {circle, draw, fill = black, minimum size = 0.3cm},empty/.style = {circle, draw, fill = white, minimum size = 0.5cm}]
            \draw (0, 0) ellipse (2 and 1);
            \node[empty,label=center:$v_1$] (A) at (-1.5,2) {};
            \node[empty,label=center:$v_4$] (D) at (1.5,2) {};
            \draw (A) -- (-1,0);
            \draw (D) -- (1,0);
            \draw (D) -- (0.5,0);
            \draw (A) -- (0,0);
            \draw (A) -- (D);
        \end{tikzpicture}
    }

    \caption{(a) the graph $G$, where $v_2$ and $v_3$ are two adjacent degree-2 vertices; (b) the graph $G'$ after deleting $v_2$ and $v_3$ and adding the edge $v_1v_4$}
    \label{rr2-3}
\end{figure}

 \begin{lemma}\label{rule3path}
     Let $G'$ be a graph obtained from $G$ by applying R\ref{3-path}, then $\alpha(G)=\alpha(G')+w(v_2)$.
 \end{lemma}
 \begin{proof}
     Let $S$ be a maximum weighted independent set in $G$. If $S$ contains $v_1$, we can assume that $S$ also contains $v_3$, because $S$ must contain one of $v_3$ and $v_4$ due to the maximality of $S$ and if $S$ contains $v_4$ we can replace $v_4$ with $v_3$ without decreasing the total weight of the independent set. For this case, $S'=S\setminus \{v_3\}$ is an independent set in $G'$ with weight $w_{G'}(S')=w_{G}(S)-w_{G}(v_2)$.
     If $S$ does not contain $v_1$, we can assume that $S$ contains $v_2$. For this case, $S'=S\setminus \{v_2\}$ is an independent set in $G'$ with weight $w_{G'}(S')=w_{G}(S)-w_{G}(v_2)$.

     On the other hand, for any maximum weighted independent set $S'$ in $G'$, we can construct a weighted independent set $S$ of $G$ such that $w_{G}(S)=w_{G'}(S')+w_{G}(v_2)$.
     If $v_1\in S'$, then $S=S'\cup\{v_3\}$ is an independent set in $G$ with weight $w_{G}(S)=w_{G'}(S')+w_{G}(v_2)$.
     If $v_1\notin S'$, then $S=S'\cup\{v_2\}$ is an independent set in $G$ with weight $w_{G}(S)=w_{G'}(S')+w_{G}(v_2)$.
 \end{proof}

\begin{reduction}[R7]
    \label{4-cycle}
    If there is a $4$-cycle $v_1v_2v_3v_4$ such that $d_G(v_2)=d_G(v_3)=2$ and $w(v_1)\geq w(v_2)\geq w(v_3)$,
    then remove $v_2$ and $v_3$, update the weight of $v_1$ by letting
    $w(v_1):=w(v_1)+w(v_3)-w(v_2)$, and add $w(v_2)$ to $M_c$.
\end{reduction}

Fig.~\ref{rr2-4} gives an illustration of R\ref{4-cycle}.  A proof of the correctness of this reduction rule is given below.

\begin{figure}
    \centering
    \subfigure{
        \centering
        \begin{tikzpicture}
            [scale = 0.7, line width = 0.5pt,solid/.style = {circle, draw, fill = black, minimum size = 0.3cm},empty/.style = {circle, draw, fill = white, minimum size = 0.5cm}]
            \draw (0, 0) ellipse (2 and 1);
            \node[empty,label=center:$v_1$] (A) at (-1.5,2) {};
            \node[empty,label=center:$v_2$] (B) at (-0.7,3) {};
            \node[empty,label=center:$v_3$] (C) at (0.7,3) {};
            \node[empty,label=center:$v_4$] (D) at (1.5,2) {};
            \draw (A) -- (-1,0);
            \draw (D) -- (1,0);
            \draw (D) -- (0.5,0);
            \draw (A) -- (0,0);
            \draw[->] (A) -- (B);
            \draw[->] (B) -- (C);
            \draw (C) -- (D);
            \draw (A) -- (D);
        \end{tikzpicture}

    }
    \qquad\qquad\qquad
    \subfigure{
        \centering
        \begin{tikzpicture}
            [scale = 0.7, line width = 0.5pt,solid/.style = {circle, draw, fill = black, minimum size = 0.3cm},empty/.style = {circle, draw, fill = white, minimum size = 0.5cm}]
            \draw (0, 0) ellipse (2 and 1);
            \node[empty,label=center:$v_1$] (A) at (-1.5,2) {};
            \node[empty,label=center:$v_4$] (D) at (1.5,2) {};
            \draw (A) -- (-1,0);
            \draw (D) -- (1,0);
            \draw (D) -- (0.5,0);
            \draw (A) -- (0,0);
            \draw (A) -- (D);
        \end{tikzpicture}

    }

    \caption{(a) the graph $G$, where $v_2$ and $v_3$ are two adjacent degree-2 vertices in a 4-cycle $v_1v_2v_3v_4$; (b) the graph $G'$ after deleting $v_2$ and $v_3$ from $G$}
    \label{rr2-4}
\end{figure}
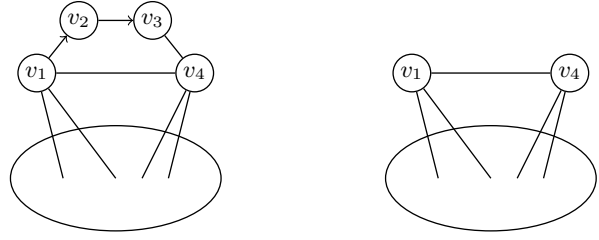

 \begin{lemma}
     Let $G'$ be a graph obtained from $G$ by applying R\ref{4-cycle}, then $\alpha(G)=\alpha(G')+w(v_2)$.
 \end{lemma}
 \begin{proof}
 This lemma can be proved analogously with the proof of Lemma~\ref{rule3path}.

     Let $S$ be a maximum weighted independent set in $G$. If $S$ contains $v_1$, we can assume that $S$ also contains $v_3$, because $S$ can not contain $v_1$ and $v_4$ now. For this case, $S'=S\setminus \{v_3\}$ is an independent set in $G'$ with weight $w_{G'}(S')=w_{G}(S)-w_{G}(v_2)$.
 If $S$ does not contain $v_1$, we can assume that $S$ contains $v_2$, since $S$ must contain at least one of $v_1$ and $v_2$ and $w(v_1)\geq w(v_2)$. For this case, $S'=S\setminus \{v_2\}$ is an independent set in $G'$ with weight $w_{G'}(S')=w_{G}(S)-w_{G}(v_2)$.

 On the other hand, for any maximum weighted independent set $S'$ in $G'$, we can construct a weighted independent set $S$ of $G$ such that $w_{G}(S)=w_{G'}(S')+w_{G}(v_2)$.
 If $v_1\in S'$, then $S=S'\cup\{v_3\}$ is an independent set in $G$ with weight $w_{G}(S)=w_{G'}(S')+w_{G}(v_2)$.
     If $v_1\notin S'$, then $S=S'\cup\{v_2\}$ is an independent set in $G$ with weight $w_{G}(S)=w_{G'}(S')+w_{G}(v_2)$.
 \end{proof}

    Next, we introduce more rules for degree-2 vertices in some complicated structures.

\begin{reduction}[R8]
    \label{4-path}
    If there is a 4-path $v_1v_2v_3v_4v_5$ such that $d_G(v_2)=d_G(v_3)=d_G(v_4)=2$ and $w(v_1)\geq w(v_2)\geq w(v_3)\leq w(v_4)\leq w(v_5)$,
    then remove $v_2$ and $v_4$, add edges $v_1v_3$ and $v_3v_5$, update the weight of $v_1$ by letting $w(v_1):=w(v_1)+w(v_3)-w(v_2)$ and the weight of $v_5$ by letting $w(v_5):=w(v_5)+w(v_3)-w(v_4)$,
    and add $w(v_2)+w(v_4)-w(v_3)$ to $M_c$.
\end{reduction}

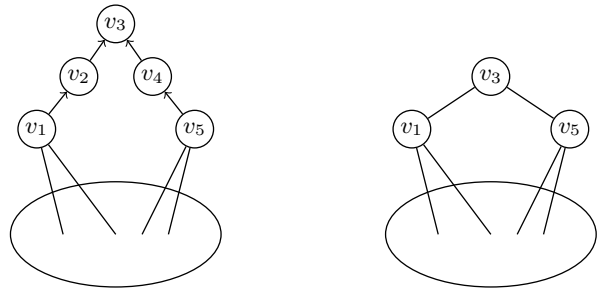
\begin{figure}
    \centering
    \subfigure{
        \centering
        \begin{tikzpicture}
            [scale = 0.7, line width = 0.5pt,solid/.style = {circle, draw, fill = black, minimum size = 0.3cm},empty/.style = {circle, draw, fill = white, minimum size = 0.5cm}]
            \draw (0, 0) ellipse (2 and 1);
            \node[empty,label=center:$v_1$] (A) at (-1.5,2) {};
            \node[empty,label=center:$v_2$] (B) at (-0.7,3) {};
            \node[empty,label=center:$v_3$] (C) at (0,4) {};
            \node[empty,label=center:$v_4$] (D) at (0.7,3) {};
            \node[empty,label=center:$v_5$] (E) at (1.5,2) {};
            \draw (A) -- (-1,0);
            \draw (E) -- (1,0);
            \draw (E) -- (0.5,0);
            \draw (A) -- (0,0);
            \draw[->] (A) -- (B);
            \draw[->] (B) -- (C);
            \draw[->] (D) -- (C);
            \draw[->] (E) -- (D);
        \end{tikzpicture}

    }
    \qquad\qquad\qquad
    \subfigure{
        \centering
        \begin{tikzpicture}
            [scale = 0.7, line width = 0.5pt,solid/.style = {circle, draw, fill = black, minimum size = 0.3cm},empty/.style = {circle, draw, fill = white, minimum size = 0.5cm}]
            \draw (0, 0) ellipse (2 and 1);
            \node[empty,label=center:$v_1$] (A) at (-1.5,2) {};
            \node[empty,label=center:$v_3$] (C) at (0,3) {};
            \node[empty,label=center:$v_5$] (E) at (1.5,2) {};
            \draw (A) -- (-1,0);
            \draw (E) -- (1,0);
            \draw (E) -- (0.5,0);
            \draw (A) -- (0,0);
            \draw (A) -- (C);
            \draw (E) -- (C);
        \end{tikzpicture}
    }

    \caption{(a) the graph $G$ with three adjacent degree-2 vertices $v_2, v_3$ and $v_4$ in a 4-path; (b) the graph $G'$ after deleting $v_2$ and $v_4$ and adding edges $v_2v_3$ and $v_3v_4$}
    \label{rr2-5}
\end{figure}

Fig.~\ref{rr2-5} gives an illustration of R\ref{4-path}. The correctness of this reduction rule is based on the following lemma.

\begin{lemma}
    Let $G'$ be a graph obtained from $G$ by applying R\ref{4-path}, then $\alpha(G)=\alpha(G')+w(v_2)+w(v_4)-w(v_3)$.
\end{lemma}
\begin{proof}
 Let $S$ be a maximum weighted independent set in $G$. We consider the following four cases.

 Case 1. $S$ contains both of $v_1$ and $v_5$: For this case, we can assume that $v_3$ is also in $S$ by the maximality of $S$.
 We can see that $S'=S\setminus \{v_3\}$ is an independent set in $G'$ with weight $w_{G'}(S')=w_{G}(S)+w_G(v_3)-w_G(v_2)+w_G(v_3)-w_G(v_4)-w_G(w_3)= w_{G}(S) +w_G(v_3)-w_G(v_2)-w_G(v_4)$.

 Case 2.  $S$ contains $v_1$ but not $v_5$: For this case, we can assume that $S$ also contains $v_4$, because $S$ must contain one of $v_3$ and $v_4$ due to the maximality of $S$ and $w(v_3)\leq w(v_4)$.
 For this case, $S'=S\setminus \{v_4\}$ is an independent set in $G'$ with weight $w_{G'}(S')= w_{G}(S) +w_G(v_3)-w_G(v_2)-w_G(v_4)$.

 Case 3.  $S$ contains $v_5$ but not $v_1$: For this case, we can assume that $S$ also contains $v_2$, because $S$ must contain one of $v_3$ and $v_2$ due to the maximality of $S$ and $w(v_3)\leq w(v_2)$.
 For this case, $S'=S\setminus \{v_2\}$ is an independent set in $G'$ with weight $w_{G'}(S')= w_{G}(S) +w_G(v_3)-w_G(v_2)-w_G(v_4)$.

 Case 4.  $S$ contains none of $v_1$ and $v_5$: For this case, we can assume that $S$  contains both of $v_2$ and $v_4$.
 For this case, $S'=S\cup\{v_3\}\setminus\{v_2,v_4\}$ is an independent set in $G'$ with weight $w_{G'}(S')= w_{G}(S) +w_G(v_3)-w_G(v_2)-w_G(v_4)$.

On the other hand, for any maximum weighted independent set $S'$ in $G'$, we can construct a weighted independent set $S$ in $G$ such that $w_{G}(S)=w_{G'}(S')+w_G(v_2)+w_G(v_4)-w_G(v_3)$.
If $S'$ does not contain any of $v_1$ and $v_5$, then it must contain $v_3$ due to the maximality of $S$. For this case, $S=S'\cup\{v_2,v_4\}\setminus\{v_3\}$ is an satisfied independent set in $G$. If $S'$ contains both of $v_1$ and $v_5$, then $S=S'\cup\{v_3\}$ is an satisfied independent set in $G$.
If $S'$ contains $v_1$ but not $v_5$, then $S=S'\cup\{v_4\}$ is an satisfied independent set in $G$. If $S'$ contains $v_5$ but not $v_1$, then $S=S'\cup\{v_2\}$ is an satisfied independent set in $G$.

\end{proof}

\begin{reduction}[R9]
    \label{5-cycle}
    For a 5-cycle $v_1v_2v_3v_4v_5$ such that $d_G(v_2)=d_G(v_3)=d_G(v_5)=2$, $\min\{d(v_1),d(v_4)\}\geq 3$, and $w(v_1)\geq w(v_2)\geq w(v_3)\leq w(v_4)$,
    \begin{enumerate}[(1)]
        \item if $w(v_3)>w(v_5)$, then remove $v_5$, update the weight of $v_i$ by letting $w(v_i):=w(v_i)-w(v_5)$ for $i=1,2,3,4$, and add $2w(v_5)$ to $M_c$.
        \item if $w(v_3)\leq w(v_5)$, then remove $v_2$ and $v_3$,  update the weight of $v_1$ by letting $w(v_1):=w(v_1)-w(v_2)$, the weight of $v_4$ by letting $w(v_4):=w(v_4)-w(v_3)$ and the weight of $v_5$ by letting $w(v_5):=w(v_5)-w(v_3)$, and add $w(v_2)+w(v_3)$ to $M_c$.
    \end{enumerate}
\end{reduction}

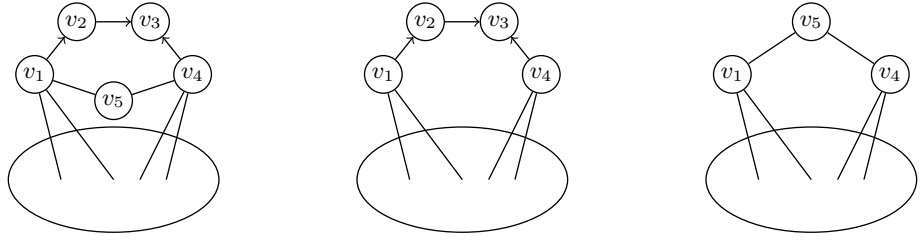
\begin{figure}
    \centering

    \subfigure{
        \centering
        \begin{tikzpicture}
            [scale = 0.7, line width = 0.5pt,solid/.style = {circle, draw, fill = black, minimum size = 0.3cm},empty/.style = {circle, draw, fill = white, minimum size = 0.5cm}]
            \draw (0, 0) ellipse (2 and 1);
            \node[empty,label=center:$v_1$] (A) at (-1.5,2) {};
            \node[empty,label=center:$v_2$] (B) at (-0.7,3) {};
            \node[empty,label=center:$v_3$] (C) at (0.7,3) {};
            \node[empty,label=center:$v_4$] (D) at (1.5,2) {};
            \node[empty,label=center:$v_5$] (E) at (0,1.5) {};
            \draw (A) -- (-1,0);
            \draw (D) -- (1,0);
            \draw (D) -- (0.5,0);
            \draw (A) -- (0,0);
            \draw[->] (A) -- (B);
            \draw[->] (B) -- (C);
            \draw[->] (D) -- (C);
            \draw (A) -- (E);
            \draw (D) -- (E);
        \end{tikzpicture}
    }
    \hfill
    \subfigure{
        \centering
        \begin{tikzpicture}
            [scale = 0.7, line width = 0.5pt,solid/.style = {circle, draw, fill = black, minimum size = 0.3cm},empty/.style = {circle, draw, fill = white, minimum size = 0.5cm}]
            \draw (0, 0) ellipse (2 and 1);
            \node[empty,label=center:$v_1$] (A) at (-1.5,2) {};
            \node[empty,label=center:$v_2$] (B) at (-0.7,3) {};
            \node[empty,label=center:$v_3$] (C) at (0.7,3) {};
            \node[empty,label=center:$v_4$] (D) at (1.5,2) {};
            \draw (A) -- (-1,0);
            \draw (D) -- (1,0);
            \draw (D) -- (0.5,0);
            \draw (A) -- (0,0);
            \draw[->] (A) -- (B);
            \draw[->] (B) -- (C);
            \draw[->] (D) -- (C);
        \end{tikzpicture}
    }
    \hfill
    \subfigure{
        \centering
        \begin{tikzpicture}
            [scale = 0.7, line width = 0.5pt,solid/.style = {circle, draw, fill = black, minimum size = 0.3cm},empty/.style = {circle, draw, fill = white, minimum size = 0.5cm}]
            \draw (0, 0) ellipse (2 and 1);
            \node[empty,label=center:$v_1$] (A) at (-1.5,2) {};
            \node[empty,label=center:$v_4$] (D) at (1.5,2) {};
            \node[empty,label=center:$v_5$] (E) at (0,3) {};
            \draw (A) -- (-1,0);
            \draw (D) -- (1,0);
            \draw (D) -- (0.5,0);
            \draw (A) -- (0,0);
            \draw (A) -- (E);
            \draw (D) -- (E);
        \end{tikzpicture}

    }
    \caption{(a) the graph $G$ containing a 5-cycle $v_1v_2v_3v_4v_5$ with three degree-2 vertices $v_2,v_3$ and $v_5$; (b) the graph $G'$ after applying (1) in R\ref{5-cycle} on $G$; (c) the graph $G'$ after applying (2) in R\ref{5-cycle} on $G$}
    \label{rr2-6}
\end{figure}

Fig.~\ref{rr2-6} gives an illustration of R\ref{5-cycle}. The correctness of this reduction rule is based on the following lemma.

\begin{lemma}
    Let $G'$ be the graph obtained from $G$ by applying R\ref{5-cycle}.
    If (1) of R\ref{5-cycle} is applied, then $$\alpha(G)=\alpha(G')+2w(v_5);$$
    If (2) of R\ref{5-cycle} is applied, then $$\alpha(G)=\alpha(G')+w(v_2)+w(v_3).$$
\end{lemma}
\begin{proof}
It is easy to observe that any maximum weighted independent set $S$ will contain exactly two non-adjacent vertices in the 5-cycle $v_1v_2v_3v_4v_5$. There are only five possible cases.
If $\{v_3,v_5\}\subseteq S$, then $S\cup\{v_2\}\setminus\{v_3\}$ is another independent set in $G$ with the weight at least $w(S)$. So we can assume that $S$ contains
one of $\{v_1,v_4\}$, $\{v_1,v_3\}$, $\{v_2,v_4\}$ and $\{v_2,v_5\}$.

Case (1): We assume that (1) of R\ref{5-cycle} is applied, where $w(v_3)> w(v_5)$.
    Let $S^*$ denote one of $\{v_1,v_4\}$, $\{v_1,v_3\}$ and $\{v_2,v_4\}$.
    If $S^*\subseteq S$, then $S'=S$ is an independent set in $G'$ such that $w_{G'}(S')=w_G(S\setminus S^*)+w_{G'}(S^*)=w_G(S)-2w_G(v_5)$.
    Otherwise, $\{v_2,v_5\}\subseteq S$.
    For this case, $S'=S\setminus\{v_5\}$ is a weighted independent set in $G'$ such that $w_{G'}(S')=w_G(S'\setminus\{v_2\})+w_{G'}(v_2)=w_G(S)-2w_G(v_5)$.

    On the other hand, for any maximum weighted independent set $S'$ in $G'$, we can construct a weighted independent set $S$ in $G$ such that $w_G(S)=w_{G'}(S')+2w_G(v_5)$.
    We can assume that $S'\cap \{v_1,v_2,v_3,v_4\}$ is one of $\{v_1,v_3\}$, $\{v_2,v_4\}$, $\{v_1,v_4\}$ and $\{v_2\}$, because if $S'$ contains $v_1$ then $S'$ also contains one of $v_3$ and $v_4$ and if
    $S'$ does not contain $v_1$ then we can assume that $S'$ contains $v_2$.

    Let $S^*$ denote one of $\{v_1,v_4\}$, $\{v_1,v_3\}$ and $\{v_2,v_4\}$.
    If $S^*\subseteq S'$, then $S=S'$ is an independent set in $G$ such that $w_G(S)=w_{G'}(S'\setminus S^*)+w_G(S^*)=w_{G'}(S')+2w_G(v_5)$.
    Otherwise, $v_2\in S'$ but $v_4\notin S'$.
    For this case, $S=S'\cup\{v_5\}$ is an independent set in $G$ such that $w_G(S)=w_{G'}(S'\setminus\{v_2\})+w_G(\{v_2,v_5\})=w_{G'}(S')+2w_G(v_5)$.

Case (2): We assume that (2) of R\ref{5-cycle} is applied, where $w(v_3)\leq w(v_5)$.
    Let $S^*$ denote one of $\{v_1,v_4\}$, $\{v_1,v_3\}$ and $\{v_2,v_4\}$.
    If $S^*\subseteq S$, then $S'=S\setminus\{v_2,v_3\}$ is an independent set in $G'$ such that $w_{G'}(S')=w_G(S\setminus (S^*\cup\{v_2,v_3\}))+w_{G'}(S^*\setminus\{v_2,v_3\})=w_G(S)-w_G(\{v_2,v_3\})$.
    Otherwise, $\{v_2,v_5\}\subseteq S$.
    For this case, $S'=S\setminus\{v_2\}$ is an independent set in $G'$ such that $w_{G'}(S')=w_G(S'\setminus\{v_5\})+w_{G'}(v_5)=w_G(S)-w_G(\{v_2,v_3\})$.

    On the other hand, for any maximum weighted independent set $S'$ in $G'$, we can construct a weighted independent set $S$ in $G$ such that $w_G(S)=w_{G'}(S')+w_G(\{v_2,v_3\})$.
    If $\{v_1,v_4\}\subseteq S'$, then $S=S'$ is an independent set in $G$ such that $w_G(S)=w_{G'}(S'\setminus\{v_1,v_4\})+w_G(\{v_1,v_4\})=w_{G'}(S')+w_G(\{v_2,v_3\})$.
    If exactly one vertex of $v_1$ and $v_4$, say $v$ is in $S'$, then $S=S'\cup(\{v_2,v_3\}\setminus N(v))$ is an independent set in $G$ such that $w_G(S)=w_{G'}(S'\setminus \{v\})+w_G(\{v\}\cup\{v_2,v_3\}\setminus N(v))=w_{G'}(S')+w_{G}(\{v_2,v_3\})$.
    Otherwise, $v_5\in S'$.
    For this case, $S=S'\cup\{v_2\}$ is an independent set in $G$ such that $w_G(S)=w_{G'}(S'\setminus\{v_5\})+w_G(\{v_2,v_5\})=w_{G'}(S')+w_G(\{v_2,v_3\})$.

\end{proof}

\begin{reduction}[R10]
    \label{6-cycle}
    For a 6-cycle $v_1v_2v_3v_4v_5v_6$ such that $d_G(v_2)=d_G(v_3)=d_G(v_5)=d_G(v_6)=2$,
    $w(v_1)\geq \max\{w(v_2),w(v_6)\}$, $w(v_4)\geq \max\{w(v_3),w(v_5)\}$, and $w(v_6)\geq w(v_5)$,
    \begin{enumerate}[(1)]
        \item if $w(v_2)\geq w(v_3)$, then remove $v_5$ and $v_6$, and update the weight of $v_2$ by letting $w(v_2):=w(v_2)+w(v_6)$ and the weight of $v_3$ by letting $w(v_3):=w(v_3)+w(v_5)$;
        \item if $w(v_2)< w(v_3)$, then remove $v_6$, add edge $v_1v_5$, and update the weight of $v_2$ by letting $w(v_2):=w(v_2)+w(v_6)$, the weight of $v_3$ by letting $w(v_3):=w(v_3)+w(v_5)$, and the weight of $v_5$ by letting $w(v_5):=w(v_6)+w(v_3)-\max\{w(v_2)+w(v_6),w(v_3)+w(v_5)\}$.
    \end{enumerate}
\end{reduction}

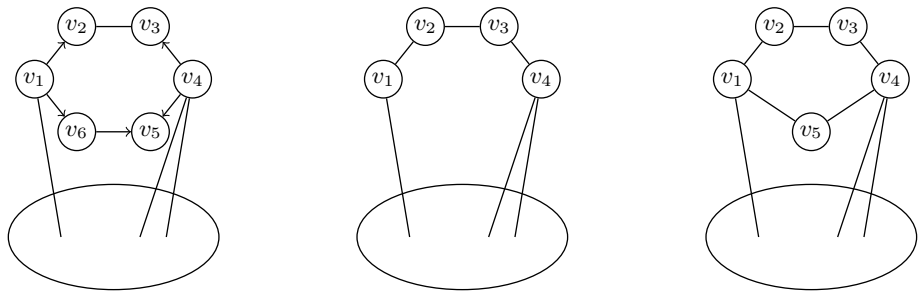
\begin{figure}
    \centering
    \subfigure{
        \centering
        \begin{tikzpicture}
            [scale = 0.7, line width = 0.5pt,solid/.style = {circle, draw, fill = black, minimum size = 0.3cm},empty/.style = {circle, draw, fill = white, minimum size = 0.5cm}]
            \draw (0, -1) ellipse (2 and 1);
            \node[empty,label=center:$v_1$] (A) at (-1.5,2) {};
            \node[empty,label=center:$v_2$] (B) at (-0.7,3) {};
            \node[empty,label=center:$v_3$] (C) at (0.7,3) {};
            \node[empty,label=center:$v_4$] (D) at (1.5,2) {};
            \node[empty,label=center:$v_5$] (E) at (0.7,1) {};
            \node[empty,label=center:$v_6$] (F) at (-0.7,1) {};
            \draw (A) -- (-1,-1);
            \draw (D) -- (1,-1);
            \draw (D) -- (0.5,-1);
            \draw[->] (A) -- (B);
            \draw[->] (A) -- (F);
            \draw[->] (D) -- (C);
            \draw[->] (D) -- (E);
            \draw (B) -- (C);
            \draw[->] (F) -- (E);
        \end{tikzpicture}

    }
    \hfill
    \subfigure{
        \centering
        \begin{tikzpicture}
            [scale = 0.7, line width = 0.5pt,solid/.style = {circle, draw, fill = black, minimum size = 0.3cm},empty/.style = {circle, draw, fill = white, minimum size = 0.5cm}]
            \draw (0, -1) ellipse (2 and 1);
            \node[empty,label=center:$v_1$] (A) at (-1.5,2) {};
            \node[empty,label=center:$v_2$] (B) at (-0.7,3) {};
            \node[empty,label=center:$v_3$] (C) at (0.7,3) {};
            \node[empty,label=center:$v_4$] (D) at (1.5,2) {};
            \draw (A) -- (-1,-1);
            \draw (D) -- (1,-1);
            \draw (D) -- (0.5,-1);
            \draw (A) -- (B);
            \draw (D) -- (C);
            \draw (B) -- (C);
        \end{tikzpicture}
    }
    \hfill
    \subfigure{
        \centering
        \begin{tikzpicture}
            [scale = 0.7, line width = 0.5pt,solid/.style = {circle, draw, fill = black, minimum size = 0.3cm},empty/.style = {circle, draw, fill = white, minimum size = 0.5cm}]
            \draw (0, -1) ellipse (2 and 1);
            \node[empty,label=center:$v_1$] (A) at (-1.5,2) {};
            \node[empty,label=center:$v_2$] (B) at (-0.7,3) {};
            \node[empty,label=center:$v_3$] (C) at (0.7,3) {};
            \node[empty,label=center:$v_4$] (D) at (1.5,2) {};
            \node[empty,label=center:$v_5$] (E) at (0,1) {};
            \draw (A) -- (-1,-1);
            \draw (D) -- (1,-1);
            \draw (D) -- (0.5,-1);
            \draw (A) -- (B);
            \draw (D) -- (C);
            \draw (D) -- (E);
            \draw (B) -- (C);
            \draw (A) -- (E);
        \end{tikzpicture}

    }

    \caption{(a) the graph $G$ containing a 6-cycle $v_1v_2v_3v_4v_5v_6$ with four degree-2 vertices $v_2,v_3,v_5$ and $v_6$; (b) the graph $G'$ after applying (1) in R\ref{6-cycle} on $G$; (c) the graph $G'$ after applying (2) in R\ref{6-cycle} on $G$}
    \label{rr2-7}
\end{figure}

Fig.~\ref{rr2-7} gives an illustration of R\ref{6-cycle}. The correctness of this reduction rule is based on the following lemma.

\begin{lemma}
    Let $G'$ be the graph obtained from $G$ by applying R\ref{6-cycle}, then $\alpha(G)=\alpha(G')$.
\end{lemma}
\begin{proof}
Case (1): We assume that (1) of R\ref{6-cycle} is applied, where $w(v_2)\geq w(v_3)$.
Let $S$ be a maximum weighted independent set in $G$.

We show that, without loss of generality, we can assume that
$S$ contains both of $v_2$ and $v_6$ or none of them (resp., both of $v_3$ and $v_5$ or none of them).
The reason is based on the following observation. If $S$ contains $v_2$ but not $v_6$, then $S$ must contain $v_5$, otherwise $v_6$ should be added to $S$ directly by the maximality of $S$. For this case, we can replace $v_5$ with $v_6$ in $S$ to get another maximum weighted independent set since $w(v_6)\geq w(v_5)$.
If $S$ contains $v_6$ but not $v_2$, then $S$ must contain $v_3$, otherwise $v_2$ should be added to $S$ directly by the maximality of $S$. For this case, we can replace $v_3$ with $v_2$ in $S$ to get another maximum weighted independent set since for this case we have that $w(v_2)\geq w(v_3)$. So we can assume that $S$ contains either both of $v_2$ and $v_6$ or none of them.
If $S$ contains $v_3$ but not $v_5$, then $S$ must contain $v_6$, otherwise $v_5$ should be added to $S$ directly by the maximality of $S$. For this case, we can replace $v_3$ with $v_2$ in $S$ to get another maximum weighted independent set since $w(v_2)\geq w(v_3)$. Now the set $S$ contains both of $v_2$ and $v_6$ but none of
$v_3$ and $v_5$. If $S$ contains $v_5$ but not $v_3$, then $S$ must contain $v_2$, otherwise $v_3$ should be added to $S$ directly by the maximality of $S$.
For this case, we can replace $v_5$ with $v_6$ in $S$ to get another maximum weighted independent set since $w(v_6)\geq w(v_5)$. We also get a maximum weighted independent set that contains both of $v_2$ and $v_6$ but none of $v_3$ and $v_5$.

If $\{v_2,v_6\}\subseteq S$, then $S'=S\setminus\{v_6\}$ is a weighted independent set in $G'$ with weight $w_{G'}(S')=w(S)$.
    If $\{v_3,v_5\}\subseteq S$, then $S'=S\setminus\{v_5\}$ is a weighted independent set in $G'$ with weight $w_{G'}(S')=w(S)$.

    On the other hand, for any maximum weighted independent set $S'$ in $G'$, we can construct a weighted independent set $S$ in $G$ such that $w_G(S)=w_{G'}(S')$.
    If $v_2\in S'$, then $S=S'\cup\{v_6\}$ is an independent set in $G$ such that $w_G(S)=w_{G'}(S')$.
    If $v_3\in S'$, then $S=S'\cup\{v_5\}$ is an independent set in $G$ such that $w_G(S)=w_{G'}(S')$.

Case (2): We assume that (2) of R\ref{6-cycle} is applied, where $w(v_2)< w(v_3)$.
Let $S$ be a maximum weighted independent set in $G$.

If $S$ contains $v_1$, then $S$ will contain either $v_4$ or
both of $v_3$ and $v_5$ (after deleting $N[v_1]$ from the graph $\{v_3,v_5\}$ will be a twin).
If $S$ does not contain $v_1$, then we can assume that $S$ contain $v_6$ since $w(v_6)\geq w(v_5)$. For this case, $S$ must contain one of $v_2$ and $v_3$ due to the maximality of $S$.
Furthermore, if $S\cap \{v_1, v_2, v_3, v_4, v_5, v_6\}=\{v_2,v_6\}$, then we can replace $v_2$ with $v_3$ in $S$ to get another maximum independent set.
So we can assume that
$S_{\Delta}=S\cap \{v_1, v_2, v_3, v_4, v_5, v_6\}$ is one of $\{v_1,v_3,v_5\}$, $\{v_1,v_4\}$,
$\{v_2,v_4,v_6\}$, and $\{v_3,v_6\}$.

   If $S_{\Delta}$ is $\{v_1,v_3,v_5\}$ or $\{v_2,v_4,v_6\}$, then $S'=S\setminus\{v_5, v_6\}$ is an independent set in $G'$ such that $w_{G'}(S')=w_G(S'\setminus\{v_2,v_3\})+w_{G'}(\{v_2,v_3\}\cap S')=w_G(S)$.
    If $S_{\Delta}= \{v_1,v_4\}$, then $S'=S$ is an independent set in $G'$ such that $w_{G'}(S')=w_G(S)$.
    Otherwise $S_{\Delta}=\{v_3,v_6\}$. For this case, $S'=(S\setminus\{v_3,v_6\})\cup\{v^*,v_5\}$ is an independent set in $G'$, where $v^*$ is the vertex in $\{v_2,v_3\}$ with the larger weight in $G'$.
     We have that $w_{G'}(S')=w_G(S\setminus\{v_3,v_6\})+w_{G'}(\{v^*,v_5\})=w_G(S)$.

    On the other hand, for any maximum weighted independent set $S'$ in $G'$, we can construct a weighted independent set $S$ in $G$ such that $w_G(S)=w_{G'}(S')$.
    In $G'$, there is a 5-cycle $v_1v_2v_3v_4v_5$ and any maximum weighted independent set in $G'$
    contains exactly two vertices in the 5-cycle. Recall that $v^*$ is the vertex in $\{v_2,v_3\}$ with the larger weight in $G'$. Let $\{v^{\star}, v^*\}=\{v_2,v_3\}$. We have that $w_{G'}(v^*)\geq w_{G'}(v^{\star})$. If $S'$ contains $\{v^{\star},v_5\}$, then we can replace $v^{\star}$ with $v^*$ in $S'$. So we can assume that $S'_{\Delta}=S'\cap \{v_1, v_2, v_3, v_4, v_5\}$ is one of the four cases $\{v_1,v_3\}$, $\{v_2,v_4\}$, $\{v_1,v_4\}$ and $\{v^*,v_5\}$.

    If $S'_{\Delta}=\{v_1,v_3\}$, then $S=S'\cup \{v_5\}$ is an independent set in $G$ such that $w_G(S)=w_G(S'\setminus\{v_1,v_3\})+w_G(\{v_1,v_3,v_5\})=w_{G'}(S')$.
    If $S'_{\Delta}=\{v_2,v_4\}$, then $S=S'\cup \{v_6\}$ is an independent set in $G$ such that $w_G(S)=w_G(S'\setminus\{v_2,v_4\})+w_G(\{v_2,v_4,v_6\})=w_{G'}(S')$.
    If $S'_{\Delta}=\{v_1,v_4\}$, then $S=S'$ is an independent set in $G$ such that $w_G(S)=w_{G'}(S')$.
    Otherwise, $S'_{\Delta}=\{v^*,v_5\}$. For this case, $S=(S'\setminus\{v^*,v_5\})\cup \{v_3,v_6\}$ is an independent set in $G$ such that $w_G(S)=w_{G'}(S'\setminus\{v,v_5\})+w_G(\{v_3,v_6\})=w_{G'}(S')$.

\end{proof}

\subsection{Reductions Based on Small Cuts}
We also have some reduction rules to deal with vertex-cuts of size one or two, which can even be used to design a polynomial-time divide-and-conquer algorithm. However, a graph may not always have vertex-cuts of small size.

\medskip
\noindent\textbf{Vertex-Cuts of Size One.} We first introduce the reduction rule based on vertex-cuts of size one.

\begin{lemma}\label{cutone}
Let $\{u\}$ be a vertex-cut of size one in $G$ and $G^*$ be a connected component in $G-u$.
\begin{enumerate}[(1)]
        \item if $w(u)+\alpha(G^*- N[u])\leq \alpha(G^*)$, then $\alpha(G)=\alpha(G_1)+\alpha(G^*)$, where $G_1$ is the remaining graph after removing $G^*$ and $u$ from $G$;
        \item if $w(u)+\alpha(G^*- N[u])> \alpha(G^*)$, then $\alpha(G)=\alpha(G_2)+\alpha(G^*)$, where $G_2$ is the remaining graph after removing $G^*$ from $G$ and updating the weight of $u$ by letting $w(u):=w(u)+\alpha(G^*- N[u])-\alpha(G^*)$.
    \end{enumerate}
\end{lemma}
\begin{proof}

    Let $G'=G-G^*-N[u]$ and $G''=G^*-N[u]$. Let $S$ be a maximum weighted independent set in $G$.

    Case (1): $\alpha(G'')+w(u)\leq \alpha(G^*)$.
    If $u\in S$, then we can see that $\alpha(G)=\alpha(G')+\alpha(G'')+w(u)$.
    Recall that we use $S(G)$ to denote a maximum independent set in $G$.
    We can replace $S\cap (\{u\}\cup V(G''))$ with $S(G^*)$ in $S$ to get another independent set
    without decreasing the total weight since $\alpha(G'')+w(u)\leq \alpha(G^*)$.
    Furthermore, we can replace $S\cap V(G')$ with $S(G_1)$ in $S$ to get another independent set in $G$.
    Thus, $w(S\cap V(G'))=\alpha(G_1)$ and then $\alpha(G)=\alpha(G_1)+\alpha(G^*)$.
    Otherwise $u\not\in S$ and it directly holds that $\alpha(G)=\alpha(G^*)+\alpha(G_1)$.

    Case (2): $\alpha(G'')+w(u)> \alpha(G^*)$.
    If $u\in S$, then $S'=S\setminus V(G^*)$ is an independent set in $G_2$ with weight $w_{G_2}(S')=w_G(S'\setminus\{u\})+w_{G_2}(u)=w_G(S'\setminus\{u\})+ w_G(u)+\alpha(G'')-\alpha(G^*)= \alpha(G)-\alpha(G^*)$.
    If $u\notin S$, then $S'=S\setminus V(G^*)$ is an independent set in $G'$ with $w_{G_2}(S')=w_G(S')=\alpha(G)-\alpha(G^*)$.

    On the other hand, for any maximum weighted independent set $S'$ in $G_2$, we can construct an independent set $S_0$ in $G$ such that $w_G(S_0)=w_{G_2}(S')+\alpha(G^*)=\alpha(G_2)+\alpha(G^*)$.
     For the case that $u\in S'$, $S_0=S'\cup S(G'')$ is an independent set in $G$. By $w_{G_2}(u)=w_G(u)+\alpha(G'')-\alpha(G^*)$, we get that $w_G(S_0)=w_G(S'\setminus\{u\})+w_G(u)+\alpha(G'')=w_{G_2}(S')+\alpha(G^*)=\alpha(G_2)+\alpha(G^*)$.
     For the case that $u\notin S'$, $S_0=S'\cup S(G^*)$ is an independent set in $G$ with $w_G(S_0)=w_{G_2}(S')+\alpha(G^*)=\alpha(G_2)+\alpha(G^*)$.

\end{proof}

\begin{figure}
    \centering
    \begin{tikzpicture}
        [scale = 0.7, line width = 0.5pt,solid/.style = {circle, draw, fill = black, minimum size = 0.3cm},empty/.style = {circle, draw, fill = white, minimum size = 0.4cm}]
        \draw (0, 2) ellipse (3 and 1);
        \draw (0,1.5) ellipse (2 and 0.5);
        \draw (0, -2) ellipse (3 and 1);
        \draw (0,-1.5) ellipse (2 and 0.5);

        \node[empty,label=center:$u$] (A) at (0,0) {};
        \node[] (C) at (-4,2) {$G^*$};
        \node[] (D) at (-5,-2) {$G-G^*-u$};
        \node[] (C1) at (0,2.5) {$G^*- N[u]$};
        \node[] (D1) at (0,-2.5) {$G-G^*-N[u]$};

        \node[empty] (E) at (1.6,1.5) {};
        \node[empty] (F) at (0.8,1.5) {};
        \node[empty] (G) at (0,1.5) {};
        \node[empty] (H) at (-0.8,1.5) {};
        \node[empty] (I) at (-1.6,1.5) {};
        \node[empty] (E1) at (1.6,-1.5) {};
        \node[empty] (F1) at (0.8,-1.5) {};
        \node[empty] (G1) at (0,-1.5) {};
        \node[empty] (H1) at (-0.8,-1.5) {};
        \node[empty] (I1) at (-1.6,-1.5) {};

        \draw (A) -- (E);
        \draw (A) -- (F);
        \draw (A) -- (G);
        \draw (A) -- (H);
        \draw (A) -- (I);

        \draw (A) -- (E1);
        \draw (A) -- (F1);
        \draw (A) -- (G1);
        \draw (A) -- (H1);
        \draw (A) -- (I1);

    \end{tikzpicture}
    \caption{A graph $G$ with a vertex-cut $\{u\}$}
    \label{f-cutone}
\end{figure}
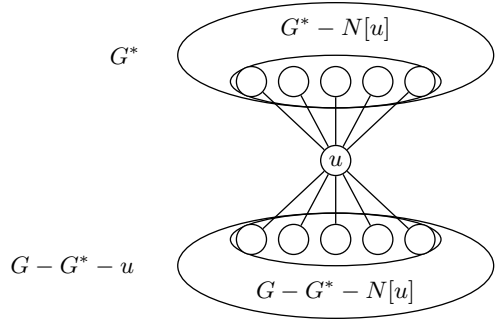

Fig.~\ref{f-cutone} shows a graph $G$ with a vertex-cut $\{u\}$.
Lemma~\ref{cutone} provides a divide-and-conquer method based on vertex-cuts of size one.
In our algorithm, we will only use it to deal with cuts that can split a connected component $G^*$ of bounded total vertex cost.

\begin{reduction}[R11]
    \label{cut_1}
     For a vertex-cut $\{u\}$ with a connected component $G^*$ in $G-u$ such that $2\delta_3-\delta_2\leq\sum_{v\in G^*}\delta_{d_G(v)} \leq 10$,
    \begin{enumerate}[(1)]
        \item if $w(u)+\alpha(G^*- N[u])\leq \alpha(G^*)$, then remove $G^*$ and $\{u\}$ from $G$ and add $\alpha(G^*)$ to $M_c$;
        \item if $w(u)+\alpha(G^*- N[u])> \alpha(G^*)$, then remove $G^*$ from $G$, update the weight of $u$ by letting $w(u):=w(u)+\alpha(G^*- N[u])-\alpha(G^*)$, and add $\alpha(G^*)$ to $M_c$.
    \end{enumerate}
\end{reduction}

\medskip
\noindent\textbf{Vertex-Cuts of Size Two.} For vertex-cuts of size two, we have a similar result. However, it will become much more complicated.

\begin{lemma}\label{cuttwo}
Let $\{u, u'\}$ be a vertex-cut of size two in $G$ and $G^*$ be a connected component in $G-\{u,u'\}$, where we assume w.l.o.g. that $\alpha(G^*-N[u])\geq \alpha(G^*-N[u'])$. We construct a new graph $G'$ from $G$ as follows: remove $G^*$; add three new vertices $\{v_1,v_2,v_3\}$ with weight $w(v_1)=\alpha(G^*-N[u'])-\alpha(G^*-N[\{u,u'\}])$, $w(v_2)=\alpha(G^*-N[u])-\alpha(G^*-N[\{u,u'\}])$ and $w(v_3)=\alpha(G^*)-\alpha(G^*-N[u])$, and add five new edges $uv_1$, $v_1v_2$, $v_2u'$, $uv_3$ and $u'v_3$. It holds that
$$ \alpha(G)=\alpha(G')+\alpha(G^*-N[\{u,u'\}]).
$$
\end{lemma}

\begin{proof}
    Let $G''=G^*-N[\{u,u'\}]$.
    Let $S$ be a maximum weight set in $G$.

    First, we show that $G'$ can be obtained from $G$ with $\alpha(G')=\alpha(G)-\alpha(G'')$.
    If $u\in S$ but $u'\notin S$, then $S'=(S\setminus V(G^*))\cup\{v_2\}$ is a weighted independent set in $G'$ with weight $w_{G'}(S')=w_G(S\setminus V(G^*))+w_{G'}(v_2)=\alpha(G)-\alpha(G'')$.
    If $u\notin S$ but $u'\in S$, then $S'=(S\setminus V(G^*))\cup\{v_1\}$ is a weighted independent set in $G'$ with weight $w_{G'}(S')=w_G(S\setminus V(G^*))+w_{G'}(v_1)=\alpha(G)-\alpha(G'')$.
    If $u\in S$ and $u'\in S$, then $S'=(S\setminus V'(G^*))$ is a weighted independent set in $G'$ with weight $w_{G'}(S')=w_G(S\setminus V(G^*))=\alpha(G)-\alpha(G'')$.
    Otherwise, $u\notin S$ and $u'\notin S$.
    For this case, we let $S'=(S\setminus V')\cup\{v_2,v_3\}$, then $S'$ is a weighted independent set in $G'$ with the weight $w_{G'}(S')=w_G(S\setminus V(G^*))+w_{G'}(\{v_2,v_3\})=\alpha(G)-\alpha(G'')$.

    On the other hand, for any maximum weighted independent set $S'$ of $G'$, we can construct a weighted independent set $S_0$ of $G$ such that $w_G(S_0)=w_{G'}(S')+\alpha(G'')=\alpha(G')+\alpha(G'')$.
    If $u\in S'$ but $u'\notin S'$, let $S_1$ be the maximum weighted independent set in $G^*-N_G[u]$, then $S_0=(S_1\cup S')\setminus\{v_1,v_2,v_3\}$ is a weighted independent set in $G$ with weight $w_G(S_0)=w_G(S'\setminus\{v_2\})+w_G(S_1)=\alpha(G')+\alpha(G'')$.
    If $u\notin S'$ but $u'\in S'$, let $S_2$ be the maximum weighted independent set in $G^*-N_G[u']$, then $S_0=(S_2\cup S')\setminus\{v_1,v_2,v_3\}$ is a weighted independent set in $G$ with weight $w_G(S_0)=w_G(S'\setminus\{v_1\})+w_G(S_3)=\alpha(G')+\alpha(G'')$.
    If $u\in S'$ but $u'\in S'$, let $S_3$ be the maximum weighted independent set in $G''$, then $S_0=(S_3\cup S')\setminus\{v_1,v_2,v_3\}$ is a weighted independent set in $G$ with weight $w_G(S_0)=w_G(S'\setminus\{v_2\})+w_G(S_1)=\alpha(G')+\alpha(G'')$.
    Otherwise, $u\notin S'$ but $u'\notin S'$.
    For this case, we let $S_4$ be the maximum weighted independent set in $G^*$, then $S_0=(S'\cup S_4)\setminus\{v_1,v_2,v_3\}$ is a weighted independent set in $G$ such that $w_G(S_0)=w_G(S'\setminus\{v_1,v_2,v_3\})+w_G(S_4)=\alpha(G')+\alpha(G'')$.

\end{proof}

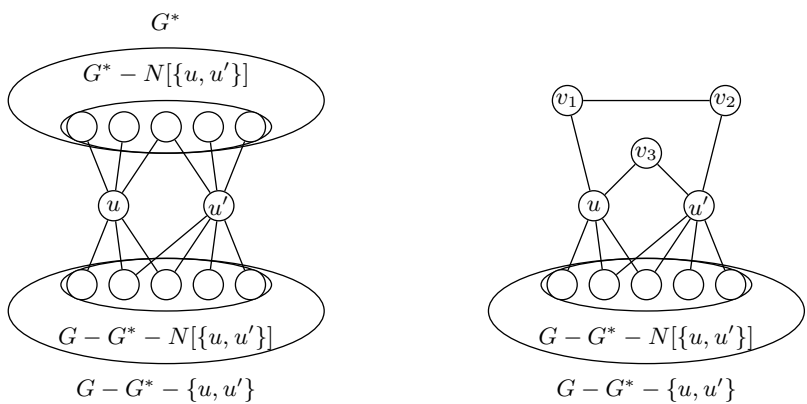
\begin{figure}
    \centering
    \subfigure{
        \centering
        \begin{tikzpicture}
            [scale = 0.7, line width = 0.5pt,solid/.style = {circle, draw, fill = black, minimum size = 0.3cm},empty/.style = {circle, draw, fill = white, minimum size = 0.4cm}]
            \draw (0, 2) ellipse (3 and 1);
            \draw (0,1.5) ellipse (2 and 0.5);
            \draw (0, -2) ellipse (3 and 1);
            \draw (0,-1.5) ellipse (2 and 0.5);

            \node[empty,label=center:$u$] (A) at (-1,0) {};
            \node[empty,label=center:$u'$] (B) at (1,0) {};
            \node[] (C) at (0,3.5) {$G^*$};
            \node[] (D) at (0,-3.5) {$G-G^*-\{u,u'\}$};
            \node[] (C1) at (0,2.5) {$G^*-N[\{u,u'\}]$};
            \node[] (D1) at (0,-2.5) {$G-G^*-N[\{u,u'\}]$};

            \node[empty] (E) at (1.6,1.5) {};
            \node[empty] (F) at (0.8,1.5) {};
            \node[empty] (G) at (0,1.5) {};
            \node[empty] (H) at (-0.8,1.5) {};
            \node[empty] (I) at (-1.6,1.5) {};
            \node[empty] (E1) at (1.6,-1.5) {};
            \node[empty] (F1) at (0.8,-1.5) {};
            \node[empty] (G1) at (0,-1.5) {};
            \node[empty] (H1) at (-0.8,-1.5) {};
            \node[empty] (I1) at (-1.6,-1.5) {};

            \draw (A) -- (G);
            \draw (A) -- (H);
            \draw (A) -- (I);

            \draw (B) -- (E);
            \draw (B) -- (F);
            \draw (B) -- (G);

            \draw (A) -- (G1);
            \draw (A) -- (H1);
            \draw (A) -- (I1);

            \draw (B) -- (E1);
            \draw (B) -- (F1);
            \draw (B) -- (G1);
            \draw (B) -- (H1);

        \end{tikzpicture}

    }
    \qquad\qquad\qquad
    \subfigure{
        \centering
        \begin{tikzpicture}
            [scale = 0.7, line width = 0.5pt,solid/.style = {circle, draw, fill = black, minimum size = 0.3cm},empty/.style = {circle, draw, fill = white, minimum size = 0.4cm}]

            \draw (0, -2) ellipse (3 and 1);
            \draw (0,-1.5) ellipse (2 and 0.5);

            \node[empty,label=center:$u$] (A) at (-1,0) {};
            \node[empty,label=center:$u'$] (B) at (1,0) {};
            \node[empty,label=center:$v_1$] (C) at (-1.5,2) {};
            \node[empty,label=center:$v_2$] (D) at (1.5,2) {};
            \node[empty,label=center:$v_3$] (E) at (0,1) {};

            \node[] (T) at (0,-3.5) {$G-G^*-\{u,u'\}$};
            \node[] (D1) at (0,-2.5) {$G-G^*-N[\{u,u'\}]$};

            \node[empty] (E1) at (1.6,-1.5) {};
            \node[empty] (F1) at (0.8,-1.5) {};
            \node[empty] (G1) at (0,-1.5) {};
            \node[empty] (H1) at (-0.8,-1.5) {};
            \node[empty] (I1) at (-1.6,-1.5) {};

            \draw (A) -- (C);
            \draw (A) -- (E);
            \draw (C) -- (D);
            \draw (D) -- (B);
            \draw (B) -- (E);

            \draw (A) -- (G1);
            \draw (A) -- (H1);
            \draw (A) -- (I1);

            \draw (B) -- (E1);
            \draw (B) -- (F1);
            \draw (B) -- (G1);
            \draw (B) -- (H1);

        \end{tikzpicture}

    }

    \caption{(a) the graph $G$ with a vertex-cut $\{u,u'\}$ of size two; (b) the graph $G'$ constructed from $G$ by removing $G^*$ and introducing three new vertices $\{v_1,v_2,v_3\}$}
    \label{fig-2-cut}
\end{figure}

 Please see Fig.~\ref{fig-2-cut} for an illustration of the construction of $G'$ in Lemma~\ref{cuttwo}.
Based on Lemma~\ref{cuttwo}, we have the following branching rule.

\begin{reduction}[R12]
    \label{cut_2}
    For a vertex-cut $\{u,u'\}$ of size two with a connected component $G^*$ in $G-\{u,u'\}$ such that $2\delta_3+\delta_2\leq\sum_{v\in G^*}\delta_{d_G(v)}\leq 10$,
    we construct the graph $G'$ in Lemma~\ref{cuttwo}, replace $G$ with $G'$, and add $\alpha(G^*-N[\{u,u'\}])$ to $M_c$.
\end{reduction}

\subsection{Analyzing Reduction Rules}

It is easy to see that all structures in our reduction rules are local structures that can be found in polynomial time. Each application of our reduction rules is to either remove some part of the graph or replace some part with a smaller structure, which can also be done in polynomial time. In this part, we mainly analyze how much the measure $p$ can be reduced in each reduction rule.
Since we will apply these reduction rules in order, we assume without loss of generality that when one reduction rule is applied, no reduction rule with a smaller index can be applied.
\begin{lemma}
    Each application of our reduction rules can be executed in polynomial time.
\end{lemma}

Each application of R\ref{remove_ucf} to R\ref{heavy_vertices} will remove some vertices from the graph. The update of the vertex weight will not increase the measure $p$.
Hence, we can get
\begin{lemma}
    \label{lrm_r1tor3}
    Each application of R\ref{remove_ucf} to R\ref{neighbor_clique}  decreases the measure $p$ by at least $\delta_{d(v)}+\sum_{u\in N(v)}\delta_{d(u)}^{<-1>}$,
    where $v$ is the vertex on which the reduction rule is applied.
\end{lemma}

\begin{lemma}
    \label{lem_triangle_c}
    Each application of R\ref{heavy_vertices} decreases the measure $p$ by at least $\sum_{u\in N[v]}\delta_{d(u)}$,
    where $v$ is the vertex on which the reduction rule is applied.
\end{lemma}

For a degree-1 vertex $v$ with the unique neighbor $u$, if $w(v)< w(u)$ then we can apply R\ref{neighbor_clique} to reduce it, otherwise $w(v)\geq w(u)$ and we can apply R\ref{heavy_vertices} to reduce it. So we know that

\begin{lemma}
    \label{triangle_no_2}
    If R\ref{remove_ucf} to R\ref{heavy_vertices} can not be applied and the graph is not empty, then the minimum degree of the graph is at least 2.
\end{lemma}

Let $C$ be a cycle that has a degree-2 vertex $v$.
If $|C|=3$, then either one of neighbors of $v$ is unconfined or $v$ can be reduced by R\ref{neighbor_clique}.
If $|C|=4$ and there is a pair of nonadjacent degree-2 vertices, then R\ref{fold_twin} can be applied.
So, we can get that

\begin{lemma}
    \label{lem_no_triangle2}
    If R\ref{remove_ucf} to R\ref{heavy_vertices} can not be applied and the graph is not empty, then there is no triangle containing degree-2 vertices and no 4-cycle containing two nonadjacent degree-2 vertices.
\end{lemma}

\begin{lemma}
    \label{lem_no_cycle_2} \label{lem_m2}
    If R\ref{remove_ucf} to R\ref{fold_d_2_2} can not be applied to $G$, then each connected component of the graph contains
    at least one vertex of degree $\geq 3$.
\end{lemma}
\begin{proof}
By Lemma~\ref{triangle_no_2}, we know that the graph has no degree-1 vertex.
If there is a connected component containing only vertices of degree $\leq 2$,
then the component can only be a cycle $C$ containing degree-2 vertices.
By Lemma~\ref{lem_no_triangle2}, we know that $|C|\geq 5$.
Let $C=\{v_1,v_2,\dots,v_{|C|}\}$.
Since R\ref{fold_d_2_2} can not be applied, we know that for each $1\leq i \leq |C|$, it holds that $w(v_i)< \max\{w(v_{i+1}),w(v_{i-1})\}$, where $v_0=v_{|C|}$ and $v_{|C|+1}=v_1$.
However, it will not hold for the vertex $v^*$ with the maximum weight in the cycle, a contradiction.
So the cycle $C$ containing only degree-2 vertices does not exist.

\end{proof}

\begin{lemma}
    \label{lem_m22}
   If R\ref{remove_ucf} to R\ref{3-path} can not be applied to $G$, then each cycle $C$ contains at least two vertices of degree $\geq 3$.
\end{lemma}
\begin{proof}
Let $C=\{v_0,v_1,v_2,\dots,v_{|C|-1}\}$ be a cycle in the graph $G$. By Lemma~\ref{lem_m2},
we know that $C$ must contain at least one vertex of degree $\geq 3$. Assume that there is exactly
one vertex of degree $\geq 3$ in $C$, which is assumed to be $v_0$ without loss of generality.
By Lemma~\ref{lem_no_triangle2}, we know that  $|C|\geq 5$.
Since R\ref{fold_d_2_2} can not be applied now, we know that $w(v_i)< \max\{w(v_{i+1}),w(v_{i-1})\}$ holds for all $1\leq i<|C|$, where $v_{|C|}=v_0$.
    Then, we can get that $w(v_0)\geq \max\{w(v_1),w(v_{|C|-1})\}$.
    Thus, either $w(v_0)\geq w(v_1)\geq w(v_2)\geq w(v_3)$ or $w(v_{|C|})\geq w(v_{|C|-1})\geq w(v_{|C|-2})\geq w(v_{|C|-3})$ holds, where implies that R\ref{3-path} can be applied.
    So $C$ must contain more than one vertices of degree $\geq 3$ and this lemma holds.

\end{proof}

\begin{lemma}
    \label{lem_5-8}
    If R\ref{remove_ucf} to R\ref{heavy_vertices} can not be applied, then each application of R\ref{fold_d_2_2} to R\ref{4-path} decreases the measure $p$ by at least $2\delta_2$.
\end{lemma}
\begin{proof}
    Assume that R\ref{remove_ucf} to R\ref{heavy_vertices} can not be applied. By Lemma~\ref{triangle_no_2}, we know that the minimum degree of $G$ is at least 2. We use $G'$ to denote the resulting graph after executing one application of R$i$ $(\ref{fold_d_2_2}\leq i \leq \ref{4-path})$.

    \textbf{Case 1:} R\ref{fold_d_2_2} is applied. A degree-2 vertex $v$ with two neighbors $u_1$ and $u_2$ are deleted from the graph, and a new vertex $v'$ with degree at most $d(u_1)+d(u_2)-2$ is introduced. For all other vertices, the degree will not increase. So the measure $p$ decreases by at least  $\delta_2+\delta_{d(u_1)}+\delta_{d(u_2)}-\delta_{d(u_1)+d(u_2)-2}$.
    Note that $\min\{d(u_1),d(u_2)\}\geq 2$. We have that $\delta_{d(u_1)}+\delta_{d(u_2)}-\delta_{d(u_1)+d(u_2)-2}\geq \delta_2+\delta_2-\delta_{2+2-2}=\delta_2$.

    \textbf{Case 2:} R\ref{3-path} is applied. We replace a chain of two degree-2 vertices with an edge. Hence, the measure $p$ decreases by $p(G)-p(G')\geq 2\delta_2$.

    \textbf{Case 3:} R\ref{4-cycle} is applied. In this case, we remove two degree-2 vertices from the graph.  Hence, the measure $p$ decreases by at least $2\delta_2$.

    \textbf{Case 4:} R\ref{4-path} is applied. In this case, two degree-2 vertices are replaced with an edge respectively. Hence, the measure $p$ decreases by $2\delta_2$.

\end{proof}

\begin{lemma}
    \label{lem_r9}
 If R\ref{remove_ucf} to R\ref{3-path} can not be applied, then each application of R\ref{5-cycle} decreases the measure $p$ by at least $2\delta_3-\delta_2$.
\end{lemma}
\begin{proof}
    Let $v_1v_2v_3v_4v_5$ be the 5-cycle that R\ref{5-cycle} is applied to, where $v_2, v_3$ and $v_5$ are degree-2 vertices, as shown in Fig.~\ref{rr2-6}.
    By Lemma~\ref{lem_m22}, we know that $\min\{d(v_1),d(v_4)\}\geq 3$.
    In this reduction rule, we will remove either a degree-2 vertex $v_5$ or two adjacent degree-2 vertices $\{v_2,v_3\}$ from the graph.
    So, we can reduce the measure by at least $\delta_2$ from the removed vertices and at
    least $2\delta_3^{<-1>}$ from the neighbors of them.

\end{proof}

\begin{lemma}
    \label{lem_r10}
 If R\ref{remove_ucf} to R\ref{3-path} can not be applied, then each application of R\ref{6-cycle}(1) decreases the measure $p$ by at least $2\delta_3$ and each application of R\ref{6-cycle}(2) decreases the measure $p$ by at least $\delta_2$.
\end{lemma}
\begin{proof}
    Let $v_1v_2v_3v_4v_5v_6$ be the 6-cycle that R\ref{6-cycle} is applied to, where $v_2, v_3, v_5$ and $v_6$ are degree-2 vertices,  as shown in Fig.~\ref{rr2-7}.
    By Lemma~\ref{lem_m22}, we know that $\min\{d(v_1),d(v_4)\}\geq 3$.
     In this reduction rule, we will either remove two adjacent degree-2 vertices $\{v_5,v_6\}$ or replace a degree-2 vertex $v_6$ with an edge.
     For the first case, the measure will decrease by at least $2\delta_3$.
     For the second case, the measure will decrease by at least $\delta_2$.

\end{proof}

\begin{lemma}
    Each application of R\ref{cut_1} decreases the measure $p$ by at least $2\delta_3-\delta_2$.
\end{lemma}
\begin{proof}
At least a subgraph $G^*$ with total cost at least $2\delta_3-\delta_2$ is deleted from the graph.
So the measure decreases by at least $2\delta_3-\delta_2$.

\end{proof}

\begin{lemma}
    \label{lem_cut2_m}
    Each application of R\ref{cut_2} will not increase the measure $p$.
\end{lemma}
\begin{proof}
    In an application of this rule, a subgraph $G^*$ will be replaced by three degree-2 vertices and the degree of the two vertices $u$ and $u'$ in the cut can increase by at most 1. So R\ref{cut_2} decreases the measure by at least $\sum_{v\in G^*}\delta_{d_G(v)}-2\delta_3-\delta_2\geq 0$.

\end{proof}
 Although an application of R\ref{cut_2} may not decrease the measure directly, it will create a 5-cycle with exactly two vertices of degree $\geq 3$, which can be further reduced by applying other reduction rules. By putting all these together, we still can strictly decrease the measure.

\begin{lemma}
    \label{lem_3d2}
     Assume that R\ref{remove_ucf} to R\ref{heavy_vertices} can not be applied. If there is a chain containing at least three degree-2 vertices, then we can apply reduction rules to decrease the measure $p$ by at least $2\delta_2$.
\end{lemma}
\begin{proof}
    Let $v_0v_1v_2v_3v_4$ be the chain such that $d(v_i)=2$ for $i=1,2,3$.
    If R\ref{fold_d_2_2} can not be applied, then we know that for $i\in{1,2,3}$, it holds that $w(v_i)\leq \max\{w(v_{i-1}),w(v_{i+1})\}$.
    So, either $w(v_0)\geq w(v_1)\geq w(v_2) \geq w(v_3)$ or $w(v_4)\geq w(v_3)\geq w(v_2) \geq w(v_1)$ or $w(v_0)\geq w(v_1)\geq w(v_2) \leq w(v_3)\leq w(v_4)$.
    Hence, one of R\ref{fold_d_2_2} to R\ref{4-path} can be applied.
    By Lemma~\ref{lem_5-8}, we know that the measure will decrease by at least $2\delta_2$.

\end{proof}

\begin{lemma}
    \label{lem_min2}
    Let $G$ be a graph where R\ref{remove_ucf} to R\ref{heavy_vertices} can not be applied.
    Let $G'$ be the graph after an application of one of R\ref{fold_d_2_2} to R\ref{4-path}.
    The minimum degree of $G'$ is at least 2.
\end{lemma}
\begin{proof}
    By Lemma~\ref{triangle_no_2}, we know that the minimum degree of $G$ is at least $2$.

    \textbf{Case 1:} R\ref{fold_d_2_2} is applied. We will delete a degree-2 vertex $v$ and its two neighbors
    $u_1$ and $u_2$, and introduce a new vertex $v'$ adjacent to all vertices in $N(N(v))$. If a degree-1 vertex $u$ is created, then the degree-1 vertex $u$ will be a vertex in $N(N(v))$. Thus, in $G$, $u$ is a degree-2 vertex, and $u$ and $v$ will form a twin. However, we have assumed that R\ref{fold_twin} can not be applied on $G$. So $G'$ can not contain a vertex of degree 1.

    \textbf{Case 2:} one of R\ref{3-path} and R\ref{4-path} is applied. Here, we will reduce a chain to a smaller one and do not decrease the degree of the two endpoints of the chain.
    Hence, the minimum degree of $G'$ is still at least 2.

    \textbf{Case 3:} R\ref{4-cycle} is applied. Let $v_1v_2v_3v_4$ be the cycle that R\ref{4-cycle} is applied to, as shown in Fig.~\ref{rr2-4}. By Lemma~\ref{lem_no_triangle2}, we know that $\min\{d(v_1),d(v_4)\}\geq 3$.
    Since we will delete $v_2$ and $v_3$ to obtain $G'$, the operation will only decrease the degree of $v_1$ and $v_2$ exactly 1 respectively.
    So, it holds that the minimum degree of $G'$ is at least 2.

\end{proof}

\begin{lemma}
    \label{lem_4cycle}
    Let $G$ be a graph where R\ref{remove_ucf} to R\ref{heavy_vertices} can not be applied. If there is a 4-cycle $C$ containing exactly two degree-2 vertices, then the measure $p$ can be decreased by at least $5\delta_2$ by applying reduction rules.
\end{lemma}
\begin{proof}
Let $C=v_1v_2v_3v_4$ be a 4-cycle in the graph.
    By Lemma~\ref{lem_no_triangle2}, we know that the two degree-2 vertices must be adjacent.
    We assume that $v_2$ and $v_3$ are degree-2 vertices.
    If R\ref{fold_d_2_2} and R\ref{3-path} can not be applied on the graph, then R\ref{4-cycle} must be applied.
    We consider the three cases.
    Let $G'$ be the resulting graph after executing one of R\ref{fold_d_2_2} to R\ref{4-cycle}. By Lemma~\ref{lem_min2}, we know that the minimum degree of $G'$ is at least 2.

    \textbf{Case 1:} R\ref{fold_d_2_2} is applied on a degree-2 vertex $v$.
    If $v\in \{v_2,v_3\}$, then the structure of $G'$ is isomorphic to the graph obtained from $G$ by deleting $\{v_2,v_3\}$.  So we have that $p(G)-p(G')\geq 2\delta_3\geq 5\delta_2$.
    Otherwise, after applying R\ref{fold_d_2_2}, the 4-cycle $C$ is still left in $G'$.
    For this case, we have that $p(G)-p(G')\geq 2\delta_2$ by Lemma~\ref{lem_5-8}.
    We can further apply reduction rules on the graph since the 4-cycle $C$ still exists.
    If R\ref{remove_ucf} to R\ref{heavy_vertices} can be applied to $G'$, then by Lemma~\ref{lrm_r1tor3} and Lemma~\ref{lem_triangle_c}, we can decrease the measure $p$ by at least $3\delta_2$ more.
    Otherwise, by induction, we know that the measure $p$ will eventually decrease by at least $5\delta_2$.

    \textbf{Case 2:} R\ref{3-path} is applied on a path $v'_1v'_2v'_3v'_4$, where $v'_2$ and $v'_3$ are degree-2 vertices. The two degree-2 vertices
    $v'_2$ and $v'_3$ will be replaced with an edge $v'_1v'_4$ if it does not exist.
    If edge $v'_1v'_4$ exists, then the reduction rule simply deletes $v'_2$ and $v'_3$ from the graph and the
    $p$ will decrease by at least $2\delta_2+2\delta_3^{<-1>}=2\delta_3\geq 5\delta_2$. If edge $v'_1v'_4$ does not exist, then
    $v'_1v'_2v'_3v'_4$ is different from $v_1v_2v_3v_4$ and the 4-cycle $C$ is still left in $G'$.
    By induction, we know that the measure $p$ will eventually decrease by at least $5\delta_2$.

    \textbf{Case 3:} R\ref{4-cycle} is applied on a path $v'_1v'_2v'_3v'_4$, where $v'_2$ and $v'_3$ are degree-2 vertices. We will remove two degree-2 vertices $\{v'_2,v'_3\}$ from $G$ to obtain $G'$.
    So, the measure $p$ will decrease by at least $2\delta_2+2\delta_3^{<-1>}=2\delta_3 \geq 5\delta_2$.

\end{proof}

\begin{lemma}
    \label{lem_5cycle_c}

    Let $G$ be a graph where R\ref{remove_ucf} to R\ref{heavy_vertices} can not be applied. If there is a cycle $C$ with $|C|\geq 5$ containing at most two vertices of degree at least 3,
    then applying reduction rules can decrease the measure $p$ by at least $2\delta_3-\delta_2$.
\end{lemma}
\begin{proof}
    By lemma~\ref{triangle_no_2}, we know that the minimum degree of $G$ is 2.
    Let $G'$ be the resulting graph after executing one of R\ref{fold_d_2_2} to R\ref{6-cycle}  on $G$ and let $G^*$ be the resulting graph after executing all the reduction rules on $G'$.
    We consider how many vertices of degree $\geq 3$ in the cycle $C$.

    \textbf{Case 1:} $C$ contains no vertex of degree $\geq 3$. By lemma~\ref{lem_m2}, we know that whole component $C$ will be reduced.
    So, $p(G)-p(G^*)\geq 5\delta_2\geq 2\delta_3-\delta_2$.

    \textbf{Case 2:} $C$ contains exactly one vertex of degree $\geq 3$.
    Since $|C|\geq 5$, there is a chain that contains at least 3 vertices of degree 2.
    By Lemma~\ref{lem_3d2}, we know that one of R\ref{fold_d_2_2} to R\ref{4-path} can be applied to some vertices of $C$ to obtained $G'$ and $p(G)-p(G')\geq 2\delta_2$.
    Since R\ref{fold_d_2_2}, R\ref{3-path} and R\ref{4-path} will reduce a chain to a smaller one with decreasing the length at most 2 and R\ref{4-cycle} will remove 2 adjacent vertices of degree 2 that are in a 4-cycle, there still exists a cycle $C'$ with at most 2 vertices of degree $\geq 3$ in $G'$ and $G'$ has a minimum degree at least 2.
    If R\ref{remove_ucf} to R\ref{heavy_vertices} can be applied to $G'$, by Lemma~\ref{lrm_r1tor3} and Lemma~\ref{lem_triangle_c}, we know that $p(G')-p(G^*)\geq 3\delta_2$.
    Otherwise, by Lemma~\ref{lem_no_triangle2}, we know that $|C'|\geq 4$.
    Now, we consider the cases of $|C'|$.
    If $|C'|=4$, then by Lemma~\ref{lem_4cycle}, we know that $p(G')-p(G^*)\geq 5\delta_2$.
    Otherwise, $|C'|\geq 5$. For this case, by induction, we know that $p(G')-p(G^*)\geq 2\delta_3-\delta_2$.

    \textbf{Case 3:} $C$ contains exactly two vertices of degree $\geq 3$. Clearly, one of R\ref{fold_d_2_2} to R\ref{6-cycle} can be applied to $G$.

    When one of R\ref{fold_d_2_2} to R\ref{4-path} is applied to $G$, the discussion is similar to the above case.

    When one of R\ref{5-cycle} to R\ref{6-cycle}(1) is applied to $G$, by Lemma~\ref{lem_r9} and Lemma~\ref{lem_r10}, we know that $p(G)-p(G')\geq 2\delta_3-\delta_2$.

    When R\ref{6-cycle}(2) is applied to $G$ to obtain $G'$, by R\ref{6-cycle}(2) and Lemma~\ref{lem_m22}, we know that there is a 5-cycle that contains at most two vertices of degree $\geq 3$ and there is no cycle containing at most one vertex of degree $\geq 3$.
    If R\ref{remove_ucf} to R\ref{heavy_vertices} can be applied to $G'$, then by Lemma~\ref{lrm_r1tor3} and Lemma~\ref{lem_triangle_c}, we know that $p(G')-p(G^*)\geq 2\delta_3-\delta_2$.
    Otherwise, by induction, we can get $p(G')-p(G^*)\geq 2\delta_3-\delta_2$.

    Since $5\delta_2\geq 2\delta_3-\delta_2$, we know that this lemma holds.

\end{proof}

\begin{definition}
    An instance is called \emph{reduced}, if none of our reduction rules can be applied.
\end{definition}

\begin{lemma}
    \label{chain_neighbor}
     In a reduced instance, any two degree-2 vertices in different chains have at most one common chain-neighbor of degree at least 3, and each cycle contains at least three vertices of degree $\geq 3$.
\end{lemma}
\begin{proof}
    Assume there are two degree-2 vertices in different chains that have two common chain-neighbors of degree at least 3. Then there is a cycle $C$ contains exactly two vertices of degree $\geq3$.
    By Lemma~\ref{lem_no_triangle2} and Lemma~\ref{lem_5cycle_c}, we know that there is no such cycle in a reduced instance.
\end{proof}

\begin{lemma}
    \label{triangle_cycle}
      For a triangle $C$ in a reduced instance, each vertex in $C$ is a vertex of degree at least 3 and it has a chain-neighbor of degree at least $3$ not in $C$.
\end{lemma}
\begin{proof}
    By Lemma~\ref{lem_no_triangle2}, we know that all the vertices in triangles are of degree at least 3.
    If a vertex in $C$ does not have a chain-neighbor of degree at least $3$ in $C$, then there is a cycle containing at most two vertices of degree $\geq 3$, a contradiction to Lemma~\ref{chain_neighbor}.

\end{proof}

\section{Branching Rules}
Next, we introduce our branching rules, which will only be applied to reduced instances.
After applying a branching rule, the algorithm will apply reduction rules as much as possible on each sub instance. In our analysis, we will consider the following applications of reduction rules together.

\subsection{Two Branching rules}
We use two branching rules. The first branching rule is to branch on a vertex $v$ by considering two cases: (i) there is a maximum weighted independent set in $G$ which does not contain $v$;
(ii) every maximum weighted independent set in $G$ contains $v$.
For the former case, we simply delete $v$ from the graph. For the latter case, by Lemma~\ref{lem_1} we know that we
can include the set $S_v$ confining $v$ in the independent set. So we delete $N[S_v]$ from the graph.

\begin{branch}[Branching on a vertex]\label{b-avertex}
    Branch on a vertex $v$ to generate two sub instances by either deleting $v$ from the graph or deleting $N[S_v]$ from the graph and adding $w(S_v)$ to $M_c$.
\end{branch}

The following property of 4-cycles has been used to design an effective branching rule for unweighted versions~\cite{DBLP:journals/tcs/XiaoN13}, which also holds in weighted graphs.

\begin{lemma}
    \label{lem_4-cycle}
    Let $v_1v_2v_3v_4$ be a cycle of length 4 in the graph $G$. Then for any independent set $S$ in $G$, either $v_1,v_3\notin S$ or $v_2,v_4\notin S$.
\end{lemma}
\begin{proof}
    Since any independent set contains at most two non-adjacent vertices in a 4-cycle, we know that this lemma  holds.

\end{proof}
Based on Lemma~\ref{lem_4-cycle}, we get the second branching rule.
\begin{branch}[Branching on a 4-cycle]
    \label{br_4cycle}
    Branch on a 4-cycle $v_1v_2v_3v_4$ to generate two sub instances by deleting either $\{v_1,v_3\}$ or
     $\{v_2,v_4\}$ from the graph.
\end{branch}

\subsection{The analysis and some properties}
The hardest part is to analyze how much we can decrease the measure in each sub-branch of a branching operation. Usually, we need to deeply analyze the local graph structure and use case-analysis.
Here we try to summarize some common properties. The following notations will be frequently used in the whole paper.

Let $S$ be a vertex subset in a reduced graph $G$. We use $G_{-S}$ to denote the graph after deleting $S$ from $G$ and iteratively applying R\ref{remove_ucf} to R\ref{heavy_vertices} until none of them can be applied. We use $R_S$ to denote the set of deleted vertices during applying R\ref{remove_ucf} to R\ref{heavy_vertices} on $G-S$.
    Then $G_{-S}=G-(S\cup R_S)$.
We also use $e_S$ to denote the number of edges between $S\cup R_S$ and $V\setminus (S\cup R_S)$ in $G$.
We have the following lemmas for some bounds on $p(G)-p(G_{-S})$.
    Note that $G_{-S}$ may not be a reduced graph because of reduction rules from R\ref{fold_d_2_2} to R\ref{cut_2} and we may further apply reduction rules to further decrease the measure $p$.
\begin{lemma}
It holds that
\color{red}
\begin{eqnarray}\label{e-tight}
p(G) - p(G_{-S})\geq \sum_{u\in S\cup R_S}\delta_{d_G(u)}+e_S \delta_3^{<-1>}.
\end{eqnarray}
 \end{lemma}
 \begin{proof}
 Since $S\cup R_S$ will be removed from $G$, we can reduce the measure by at least $\sum_{u\in S\cup R_S}\delta_{d_G(u)}$ from them. For each vertex in $N(S\cup R_S)$, it must be a vertex of degree $\geq 2$ in $G_{-S}$, otherwise R\ref{remove_ucf} to R\ref{heavy_vertices} can be further applied on  $G_{-S}$.  Thus, deleting each edge between $S\cup R_S$ and $V\setminus (S\cup R_S)$ will decrease the measure $p$ by $\delta_3^{<-1>}$ from $N(S\cup R_S)$.
In total, we can decrease the measure $p$ by $e_S\delta_3^{<-1>}$ from $N(S\cup R_S)$.

\end{proof}

In some cases, we can not use the bound in (\ref{e-tight}) directly, since we may not know the vertex set $R_S$. So we also consider some special cases and relaxed bounds.

\begin{lemma}\label{branch_1 vertex}
    Let $S=\{v\}$ be a set of a single vertex of degree $\geq 3$. We have that
    $$p(G)-p(G_{-S})\geq \delta_{d(v)}+\sum_{u\in N(v)}\delta_{d(u)}^{<-1>}+q_2\delta_{3}^{<-1>},$$
    where $q_2$ is the number of degree-2 vertices in $N(v)$.
\end{lemma}
\begin{proof}
After removing $v$, we can reduce the measure $p$ by $\delta_{d(v)}$ from the vertex $v$ itself and $\sum_{u\in N(v)}\delta_{d(u)}^{<-1>}$ from $N(v)$ since the degree of each vertex in $N(v)$ will decrease by 1.
Furthermore, each degree-2 vertex $w\in N(v)$ in $G$ will become a degree-1 vertex in $G-v$.
By Lemma~\ref{chain_neighbor}, we know that $w$ has two different chain-neighbors of degree $\geq 3$. Let they be $v$ and $v'$.
By Lemma~\ref{lem_no_triangle2}, Lemma~\ref{lem_m22}, Lemma~\ref{lem_4cycle} and Lemma~\ref{lem_5cycle_c}, we know that there is no cycle that contains at most two vertices of degree $\geq 3$.
So, we know that $v'\notin N(v)$. Furthermore, it holds that $d_{G-v}(v')=d_G(v')$.
After iteratively applying R\ref{remove_ucf} to R\ref{heavy_vertices} in $G-v$ to reduce degree-1
vertices, either $v'$ will be deleted or the degree of $v'$ will decrease by 1. Thus, we can further reduce
$p$ by at least $\min\{\delta_{d(v')},\delta_{d(v')}^{<-1>}\}=\delta_{d(v')}^{<-1>}=\delta_{3}^{<-1>}$.
By Lemma~\ref{chain_neighbor}, we know that for each degree-2 vertex $w\in N(v)$ in $G$, the other chain-neighbor $v'$ of degree $\geq 3$ is different. Thus, we can further reduce $p$ by at least $q_2\delta_{3}^{<-1>}$ from these vertices.
\end{proof}

\begin{lemma}\label{branch_1 vertexN}
If $S\cup R_S$ contains  $N[v]$ for some vertex $v$ of degree $\geq 3$, then we have that
    $$p(G)-p(G_{-S})\geq \sum_{u\in N[v]}\delta_{d(u)}+q_2\delta_{3}^{<-1>},$$
    where $q_2$ is the number of degree-2 vertices in $N(v)$.
\end{lemma}
\begin{proof}
The proof is similar to the proof of Lemma~\ref{branch_1 vertex}.
At least $N[v]\subseteq S$ is removed and we can reduce the measure $p$ by $\sum_{u\in N[v]}\delta_{d(u)}$ from $N[v]$ directly.
Each degree-2 vertex $w\in N(v)$ in $G$ has a different chain-neighbor $v'\neq v$ of degree $\geq 3$ by Lemma~\ref{chain_neighbor}, which
will be deleted or the degree of $v'$ will decrease by at least 1 after iteratively applying R\ref{remove_ucf} to R\ref{heavy_vertices} to reduce degree-1 vertices. Thus, we can further reduce $p$ by at least $q_2\delta_{3}^{<-1>}$.

\end{proof}

Recall that we use $\mathcal{C}(G')$ to denote the set of connected components of the graph $G'$. We can easily observe
the following lemma, which will be used to prove several bounds on $p(G) - p(G_{-S})$.

\begin{lemma}
    \label{lem_eq}
    Let $S$ be a vertex subset. Let $S'$ be a subset of $S\cup R_S$ and $R'=S\cup R_S\setminus S'$.
    The number of edges between $S\cup R_S$ and $V\setminus(S\cup R_S)$ is $e_S$, and the number of edges between $S'$ and $V\setminus S'$ is $k$. For any component $H\in \mathcal{C}(G[R'])$, the
    number of edges between $S'$ and $H$ is $l_H$ and the number of edges between $H$ and $N(S\cup R_S)$ is $r_H$. We have that
    $$k-e_S=\sum_{H\in \mathcal{C}(G[R'])}(l_H-r_H).$$
    Furthermore, for any component $H\in \mathcal{C}(G[R'])$ containing only degree-2 vertices, it holds that $$l_H-r_H=0~~\mbox{or}~~2.$$
\end{lemma}

 \begin{lemma}
    \label{lem_-S}
 For any subset $S'\subseteq S\cup R_S$ with $k$ edges between $S'$ and $V\setminus S'$,
  it holds that
 $$p(G) - p(G_{-S}) \geq \sum_{u\in S'}\delta_{d_G(u)} + e_S \delta_3^{<-1>} +
    \left\{
      \begin{array}{ll}
        0, & \hspace{0.5cm} k - e_S \leq 0 \\
        \delta_3, & \hspace{0.5cm} k - e_S = 1\\
        \delta_2, & \hspace{0.5cm} k - e_S = 2 \\
        \delta_3, & \hspace{0.5cm} k - e_S = 3\\
        2\delta_2, & \hspace{0.5cm} k - e_S > 3.
      \end{array}
    \right.\label{ineq_Ts_k}$$
  \end{lemma}

\begin{proof}
    By Inequality~(\ref{e-tight}), we know that this lemma holds for $k-e_S\leq 0$.
    Next, we consider the cases where $k-e_S\geq1$.
    Let $R'=S\cup R_S\setminus S'$.
    If $k-e_S=1$, then by Lemma~\ref{lem_eq}, we know that there is a component $H\in \mathcal{C}(G[R'])$ containing at least one vertex of degree $\geq 3$.
    If $k-e_S=2$, then $R'$ is nonempty and contains at least one vertex of degree $\geq 2$.
    If $k-e_S=3$, then $R'$ contains at least one vertex of degree $\geq 3$ similar to the case of $k-e_S=1$.
    If $k-e_S>3$, then $R'$ contains at least one vertex of degree $\geq 3$ or at least two vertices of degree 2.
    So, by Inequality~(\ref{e-tight}), this lemma holds.

\end{proof}

\begin{corollary}
    \label{cor_imp}
 For any subset $S'\subseteq S\cup R_S$  with $k$ edges between $S'$ and $V\setminus S'$, it holds that either
 $p(G) - p(G_{-S})> 10$ or
 $$p(G) - p(G_{-S}) \geq \sum_{u\in S'}\delta_{d_G(u)}  +
    \left\{
      \begin{array}{ll}
        k \delta_3^{<-1>}, & \hspace{0.5cm} \mbox{if}~k\leq 4 \\
         \delta_2+3\delta_3^{<-1>}, & \hspace{0.5cm} \mbox{if}~k=5 \\
         \delta_3+3\delta_3^{<-1>}, & \hspace{0.5cm} \mbox{if}~k= 6\\
         2\delta_2+3\delta_3^{<-1>}, & \hspace{0.5cm} \mbox{if}~k> 6.
      \end{array}
    \right.$$
\end{corollary}
\begin{proof}
    The instance $G$ is reduced. So R\ref{cut_1} and R\ref{cut_2} can not be applied.
    If $\sum_{v\in S\cup R_S}\delta_{d_G(v)}\leq 10$ and there are at most two edges between $S\cup R_S$ and $V\setminus (S\cup R_S)$, then the condition in R\ref{cut_1} or R\ref{cut_2} would hold.
    So we have either $p(G) - p(G_{-S})> 10$ or $e_S\geq 3$.
For the latter case, by Lemma~\ref{lem_-S}, this corollary also holds.

\end{proof}

\begin{lemma}
    \label{branch_S2}
    Assume that a reduced graph $G$ has a maximum degree 3 and has no 3 or 4-cycles.
    For any subset $S'\subseteq S\cup R_S$ with $k$ edges between $S'$ and $V\setminus S'$,
    if the diameter of the induced graph $G[S']$ is 2, then it holds that either
 $p(G) - p(G_{-S})> 10$ or
    $$p(G)-p(G_{-S})\geq \sum_{u\in S'}\delta_{d_G(u)}+3\delta_3^{<-1>}
      + \left\{
        \begin{array}{lr}
          0, \hspace{0.5cm} & k \leq 3 \\
          \delta_3^{<-1>}, \hspace{0.5cm} & k = 4 \\
          2\delta_2, \hspace{0.5cm} & k =5 \\
          \delta_2 + \delta_3, \hspace{0.5cm} & k =6.
        \end{array}
      \right.$$

\end{lemma}

\begin{proof}
    Let $R'=(S\cup R_S)\setminus S'$.
    Note that any vertex $v\in R'$ is adjacent to at most one vertex in $S'$, otherwise
    $v$ and some vertices in $S'$ would form a cycle of length at most 4 since the diameter of $G[S']$ is 2, a contradiction to the assumption that there is no cycle of length at most 4.
    Thus, any component $H\in \mathcal{C}(G[R'])$ contains at least $l_H$ vertices, and $R'$ contains at least $k-e_S$ vertices.

    We first consider the value of $k-e_S$.

    If $k-e_S=1$, then by Lemma~\ref{lem_eq}, we know that there is a vertex of degree $\geq 3$ in $R'$.
    So, by Inequality~(\ref{e-tight}), we know that
    \begin{eqnarray}\label{e-pp-1}
    p(G) - p(G_{-S}) \geq \sum_{u\in S'}\delta_{d_G(u)} + e_S\delta_3^{<-1>}+\delta_3.
    \end{eqnarray}

    If $k-e_S=2$, then $R'$ contains at least $k\geq 2$ vertices.
    So, by Inequality~(\ref{e-tight}), we know that
    \begin{eqnarray}\label{e-pp-2}
    p(G) - p(G_{-S}) \geq \sum_{u\in S'}\delta_{d_G(u)} + e_S\delta_3^{<-1>}+2\delta_2.
\end{eqnarray}

    If  $k-e_S=3$,  then $R'$ contains at least $k\geq 3$ vertices. By Lemma~\ref{lem_eq}, we know that
    at least one vertex in $R'$ is of degree $\geq 3$.
    Hence, by Inequality~(\ref{e-tight}), we know that
    \begin{eqnarray}\label{e-pp-3}
    p(G) - p(G_{-S}) \geq \sum_{u\in S'}\delta_{d_G(u)} + e_S\delta_3^{<-1>}+2\delta_2+\delta_3.
\end{eqnarray}

    Next, we prove the lemma by considering the value of $k$.

    \textbf{Case 1}: $k\leq 3$. This lemma holds by Corollary~\ref{cor_imp}.

    \textbf{Case 2}: $k=4$. If $k-e_S=0$, by Lemma~\ref{lem_-S}, we get that
    $$p(G) - p(G_{-S}) \geq \sum_{u\in S'}\delta_{d_G(u)} + 4\delta_3^{<-1>}.$$
    If $k-e_S=1,2$ and $3$, then we will get (\ref{e-pp-1}), (\ref{e-pp-2}), and (\ref{e-pp-3}), respectively.
    It is impossible that $e_S=0$ and then it is impossible that $k-e_S=4$. The worst case is that $k-e_S=0$.

    \textbf{Case 2}: $k=5$. Similar to Case 2, we consider all possible values of $k-e_S$. The
     worst case is that $k-e_S=2$, where by (\ref{e-pp-2}) we get that
    $$p(G) - p(G_{-S}) \geq \sum_{u\in S'}\delta_{d_G(u)} + 3\delta_3^{<-1>}+2\delta_2.$$

    \textbf{Case 3}: $k=6$. Similar to Case 2, we consider all possible values of $k-e_S$. The
     worst case is that $k-e_S=2$, where by (\ref{e-pp-3}) we get that
    $$p(G) - p(G_{-S}) \geq \sum_{u\in S'}\delta_{d_G(u)} + 4\delta_3^{<-1>}+2\delta_2.$$

\end{proof}

\begin{lemma}
    \label{branch_S3}
    Assume that a reduced graph $G$ has a maximum degree 3, and each cycle $C$ in it contains at least five vertices, where at least four vertices are degree-3 vertices.
    For any subset $S'\subseteq S\cup R_S$ with $k$ edges between $S'$ and $V\setminus S'$, if each path $P$ in the induced graph $G[S']$ contains either at most three vertices or at most two degree-3 vertices, then it holds that either
 $p(G) - p(G_{-S})> 10$ or

    $$p(G)-p(G_{-S})\geq\sum_{u\in S'}\delta_{d_G(u)}
    + \left\{
      \begin{array}{lr}
        k\delta_3^{<-1>},\hspace{0.5cm} & k \leq 5 \\
        \delta_3 +  2\delta_2 + 3\delta_3^{<-1>},\hspace{0.5cm} &k = 6.
      \end{array}
    \right.$$
\end{lemma}

\begin{proof}
The proof is similar to the proof of Lemma~\ref{branch_S2}. Let $R'=S\cup R_S\setminus S'$.
Any vertex $v\in R'$ is adjacent to at most one vertex in $S'$, otherwise
    $v$ would be in a 4-cycle or a cycle containing at most three degree-3 vertices since each path $P$ in $G[S']$ either contains at most three vertices or contains at most two degree-3 vertices. However, these cycles would not appear by the assumption.
    Thus, $R'$ contains at least $k-e_S$ vertices.

    We first consider the value of $k-e_S$.

    If $k-e_S=1$, then by Lemma~\ref{lem_eq}, we know that there is a vertex of degree $\geq 3$ in $R'$.
    So, by Inequality~(\ref{e-tight}), we know that
    \begin{eqnarray}\label{e-pp-4}
    p(G) - p(G_{-S}) \geq \sum_{u\in S'}\delta_{d_G(u)} + e_S\delta_3^{<-1>}+\delta_3.
    \end{eqnarray}

    If $k-e_S=2$, then $R'$ will contain at least two vertices of degree $3$. The reason is below. If
    $R'$ contains at most one degree $3$, then there will be a cycle containing at most three degree-3 vertices,
a contradiction to the assumption.
    So, by Inequality~(\ref{e-tight}), we know that
    \begin{eqnarray}\label{e-pp-5}
    p(G) - p(G_{-S}) \geq \sum_{u\in S'}\delta_{d_G(u)} + e_S\delta_3^{<-1>}+2\delta_3.
    \end{eqnarray}

   If  $k-e_S=3$,  then $R'$ contains at least $k\geq 3$ vertices. By Lemma~\ref{lem_eq}, we know that
    at least one vertex in $R'$ is of degree $\geq 3$.
    Hence, by Inequality~(\ref{e-tight}), we know that
    \begin{eqnarray}\label{e-pp-6}
    p(G) - p(G_{-S}) \geq \sum_{u\in S'}\delta_{d_G(u)} + e_S\delta_3^{<-1>}+2\delta_2+\delta_3.
\end{eqnarray}

Next, we prove the lemma by considering the value of $k$.

    \textbf{Case 1}: $k\leq 4$. This lemma holds by Corollary~\ref{cor_imp}.

    \textbf{Case 2}: $k=5$. If $k-e_S=0$, by Lemma~\ref{lem_-S}, we get that
    $$p(G) - p(G_{-S}) \geq \sum_{u\in S'}\delta_{d_G(u)} + 5\delta_3^{<-1>}.$$
    If $k-e_S=1,2$ and $3$, then we will get (\ref{e-pp-4}), (\ref{e-pp-5}), and (\ref{e-pp-6}), respectively.
    It is impossible that $e_S\leq 1$ since $G$ has no cut of size at most 2, and then it is impossible that $k-e_S\geq 4$. The worst case is that $k-e_S=0$.

    \textbf{Case 3}: $k=6$. Similar to Case 2, we consider all possible values of $k-e_S$. The
     worst case is that $k-e_S=3$, where by (\ref{e-pp-2}) we get that
   $$p(G) - p(G_{-S}) \geq \sum_{u\in S'}\delta_{d_G(u)} + 3\delta_3^{<-1>}+2\delta_2+\delta_3.$$

\end{proof}

\section{The Algorithm}
Now we are ready to describe the whole algorithm.
    The algorithm will first apply reduction rules and also solve connected components of size bounded by a constant (or the measure is bounded by a constant) directly.

Second, the algorithm will branch on vertices
of degree $\geq 5$ if any. Note that even the input graph has no high-degree vertices, some
reduction rules may create them during the algorithm. Third, the algorithm will deal with 4-cycles and degree-4 vertices. Last is to deal with degree-3 vertices, which is the most complicated part of the algorithm.
When the algorithm executes one step, we assume that all previous steps can not be applied now.

\begin{step}[Applying Reductions]
    \label{all_reductions}
    If the instance is not reduced, iteratively apply reduction rules in order, i.e., when one reduction rule is applied, no reduction rule with a smaller index can be applied on the graph.
\end{step}

\begin{step}[Solving Small Components]
    \label{small_component}
    If there is a connected component $G^*$ of $G$ such that $p(G^*)\leq 10$, solve the component $G^*$ directly and return $\alpha (G-G^*)+\alpha(G^*)$.
\end{step}

Note that $p(G^*)\leq 10$, the number of vertices with degree $\geq 3$ is at most 10. We can enumerate all subsets of vertices with degree $\geq 3$ and let them in the independent set, and the remaining graph has a
maximum degree at most 2, which can be solved in polynomial time by Lemma~\ref{lem_m2}. So this step can be solved in polynomial time.

\begin{step}[Branching on Vertices of Degree $\geq 5$]
    \label{degree_5}
    If there is a vertex $v$ with degree $d(v)\geq 5$, then branch on $v$ with Branching Rule 1 by
    either excluding $v$ from the independent set or including $S_v$ in the independent set.
\end{step}

\begin{lemma}
    Step~\ref{degree_5} followed by applications of reduction rules creates a branching
    vector covered by
\begin{eqnarray} \label{e_branching5d}
[\delta_{d(v)}+d(v)\delta_3^{<-1>}, \delta_{d(v)}+ d(v)\delta_3].
\end{eqnarray}
\end{lemma}
\begin{proof}
For the first branching, a single vertex $v$ is deleted from the graph. By Lemma~\ref{branch_1 vertex},
we get $\delta_{d(v)}+d(v)\delta_3^{<-1>}+q_2\delta_2\geq \delta_{d(v)}+d(v)\delta_3^{<-1>}$,
where $q_2\geq 0$ is the number of degree-2 vertices in $N(v)$.

For the second branching, $N[v]\subseteq N[S_v]$ is deleted from the graph.
By Lemma~\ref{branch_1 vertexN}, we know that the measure will be decreased by at least
$\sum_{u\in N[v]}\delta_{d(u)}+q_2\delta_3^{<-1>}$, which is at least $\delta_{d(v)}+ d(v)\delta_3$.

\end{proof}

For the worst case that $d(v)=5$, the branching vector (\ref{e_branching5d}) will become
$$
[\delta_{5}+5\delta_3^{<-1>}, \delta_{5}+ 5\delta_3]=[5.368,7.248].
$$
Next, we assume that the maximum degree of the graph is at most 4.

\begin{step}[Branching on 4-Cycles with Chords]
    \label{C4_with_chords}
    If there is a 4-cycle $C=v_1v_2v_3v_4$ with a chord $v_1v_3\in E$, then branch on the 4-cycle with Branching Rule~\ref{br_4cycle} by excluding either $\{v_1,v_3\}$ or $\{v_2,v_4\}$ from the independent set.
\end{step}

Note that it is impossible that both of $v_1$ and $v_3$ are of degree $3$, since otherwise one of $v_1$ and $v_3$ is an unconfined vertex and R\ref{remove_ucf} should be applied.
Then one of $v_1$ and $v_3$ is of degree 4. Without loss of generality, we assume that $v_1$ is a degree-4 vertex.
Since $v_1$ and $v_3$ are adjacent, we know that none of $v_2$ and $v_4$ can be a degree-2 vertex, otherwise R\ref{remove_ucf} to R\ref{heavy_vertices} can be applied.

\begin{lemma}
    Step~\ref{C4_with_chords} followed by applications of reduction rules creates a branching
    vector covered by one of
    $$[3\delta_4+\delta_3^{<-1>}, 4\delta_4+ 2\delta_3^{<-1>}]=[5.496,7.744]~~\mbox{and}$$
$$[4\delta_4, 2 \delta_4+2 \delta_3+2 \delta_2]=[6.496,6].$$

\end{lemma}
\begin{proof}
We use $\Delta_1$ and $\Delta_2$ to denote the amount of measure decreased in the two sub branches. We analyze
$\Delta_1$ and $\Delta_2$ by considering several different cases.

    \textbf{Case 1}: $v_2v_4\in E$. Now $\{v_1,v_2,v_3,v_4\}$ from a clique. None of them can be a degree-3 vertex in $G$, otherwise the vertex would have a clique neighborhood and R\ref{cliqueneighbor} should be applied.
    Thus, all of the four vertices in $C$ are degree-4 vertices. For each vertex in the cycle $C$, a neighbor not in $C$ is called an \emph{out-neighbor}. We can see that the out-neighbor of $v_1$ is different from the out-neighbor of $v_3$ (resp., the out-neighbor of $v_2$ is different from the out-neighbor of $v_4$), otherwise one of $v_1$ and $v_3$ (resp., one of $v_2$ and $v_4$) would be unconfined.

    In the first branching, $\{v_1,v_3\}$ is deleted from the graph.
    We can reduce the measure $p$ by $2\delta_4$ from $\{v_1,v_3\}$, $2\delta_4^{<-2>}$ from $\{v_2,v_4\}$,
    and at least $2\min \{\delta_2^{<-1>},$ $\delta_3^{<-1>},  \delta_4^{<-1>}\}=2 \delta_2$ from the out-neighbors of $\{v_1,v_3\}$. In total, it is at least $2\delta_4+2\delta_4^{<-2>}+2 \delta_2=4\delta_4$.
    In the second branching, the measure can be reduced by the same amount. We get a branching vector:
    $$
    [4\delta_4, 4\delta_4].
    $$

    Next, we assume that $v_2v_4\not\in E$.

     \textbf{Case 2}: $d(v_3)= 3$ and at least one of $v_2$ and $v_4$, say $v_2$ is a degree-3 vertex.
     In the first branching, after deleting $\{v_1,v_3\}$, vertex $v_2$ will become a degree-1 vertex and will be removed by applying R\ref{remove_ucf} to R\ref{heavy_vertices}. Thus, by Corollary~\ref{cor_imp} (setting $S'=\{v_1,v_2,v_3\}$ with $k=4$), we know that the measure $p$ decreases by at least $\sum_{u\in\{v_1,v_2,v_3\}}\delta_{d_u} + 4\delta_3^{<-1>}=2\delta_3+\delta_4+4\delta_3^{<-1>}=4\delta_4-\delta_2$.

    In the second branching, all the four vertices in the cycle $C$ will be deleted after applying R\ref{remove_ucf} to R\ref{heavy_vertices}.
    By  Corollary~\ref{cor_imp} (setting $S'=\{v_1,v_2,v_3,v_4\}$ with $k\geq 3$), we know that  the measure $p$ decreases by at least $\sum_{u\in\{v_1,v_2,v_3,v_4\}}\delta_{d_u} + 3\delta_3^{<-1>}=3\delta_3 + \delta_4+ 3\delta_3^{<-1>}=4\delta_4$. We get a branching vector
    $$
    [4\delta_4-\delta_2, 4\delta_4].
    $$

     \textbf{Case 3}: $d(v_3)= 3$ and $d(v_2)= d(v_4)= 4$.
     In the first branching of deleting $\{v_1,v_3\}$, we can reduce the measure $p$ by $\delta_4+\delta_3$
     from $\{v_1, v_3\}$, $2\delta_4^{<-2>}$ from $\{v_2, v_4\}$, and at least $\delta_2$ from the fourth neighbor of $v_1$. In total, we get $3\delta_4+\delta_3^{<-1>}$.

     The second branching is similar to the second branching in Case~2. all the four vertices in the cycle $C$ will be deleted in this branching. By  Corollary~\ref{cor_imp} (setting $S'=\{v_1,v_2,v_3,v_4\}$ with $k\geq 3$), we know that  the measure $p$ decreases by at least $\sum_{u\in\{v_1,v_2,v_3,v_4\}}\delta_{d_u} + 3\delta_3^{<-1>}=4\delta_4+ 2\delta_3^{<-1>}$. We get a branching vector
    $$
    [3\delta_4+\delta_3^{<-1>}, 4\delta_4+ 2\delta_3^{<-1>}].
    $$

\textbf{Case 4}: $d(v_3)= 4$. We show that in the first branching of deleting $\{v_1,v_3\}$, the measure $p$
will decrease by at least $4\delta_4$. If both of $v_2$ and $v_4$ are degree-4 vertices, we reduce the measure $p$ by $2 \delta_4$ from $\{v_1,v_3\}$, $2 \delta_4^{<-2>}$ from $\{v_2,v_4\}$, and at least $2 \delta_2$ from the two different out-neighbors of $\{v_1,v_3\}$.
In total, it is $2 \delta_4 + 2 \delta_4^{<-2>}+2\delta_2=4\delta_4$. Else at least one of $v_2$ and $v_4$, say $v_2$ is a degree-3 vertex. After deleting $\{v_1,v_3\}$, vertex $v_2$ will become a degree-1 vertex and will be removed by applying R\ref{remove_ucf} to R\ref{heavy_vertices}. Thus, by  Corollary~\ref{cor_imp} (setting $S'=\{v_1,v_2,v_3\}$ with $k=5$), we know that the measure $p$ decreases by at least $\sum_{u\in\{v_1,v_2,v_3\}}\delta_{d_u} + \delta_3+3\delta_3^{<-1>}=4\delta_4+\delta_3^{<-1>}> 4\delta_4$.

    In the second branching of deleting $\{v_2,v_4\}$, we reduce the measure $p$ by at least $2 \delta_3$ from $\{v_2,v_4\}$, $2 \delta_4^{<-2>}$ from $\{v_1,v_3\}$, and at least $2 \delta_2$ from two different out-neighbors of $\{v_2,v_4\}$. In total, it is $2 \delta_4+2 \delta_3+2 \delta_2$.
 We get a branching vector
    $$
    [4\delta_4, 2 \delta_4+2 \delta_3+2 \delta_2].
    $$

Note that $2 \delta_4+2 \delta_3+2 \delta_2< 4\delta_4-\delta_2< 4\delta_4$. We know that all branching vectors
will be covered by one of $[3\delta_4+\delta_3^{<-1>}, 4\delta_4+ 2\delta_3^{<-1>}]$ and
$[4\delta_4, 2 \delta_4+2 \delta_3+2 \delta_2]$.

\end{proof}

\begin{step}[Branching on Degree-4 Vertices]
    \label{S_degree4}
    If there is a degree-4 vertex $v$, then branch on it with Branching Rule 1 by
    either excluding $v$ from the independent set or including $S_v$ in the independent set.
\end{step}
We use $q_i$ to denote the number of degree-$i$ neighbors of $v$ in $G$.
In the first branching of deleting $v$, by Lemma~\ref{branch_1 vertex}, we can reduce the measure $p$ by at least
 $$\delta_4+\sum_{i=2}^4q_i\delta_i^{<-1>}+q_2\delta_3^{<-1>}\geq \delta_4+q_2\delta_3+(1-q_2)\delta_3^{<-1>}.$$

 We consider the second branching of deleting $N[S_v]$. We can see that $N[v]\subseteq N[S_v]$ will be deleted.
 Note that each degree-2 vertex in $N(v)$ is adjacent to a vertex not in $N[v]$ otherwise $v$ would have a clique neighborhood and reduction rules can be applied.
 For the case that $q_2=4$, by Lemma~\ref{branch_1 vertexN}, we know that the measure $p$ will decrease by
 at least $\delta_4+4\delta_2+4\delta_3^{<-1>}=\delta_4+4\delta_3$.
 For the case that $0\leq q_2\leq 3$, the number of edges between $N[v]$ and $V\setminus N[v]$ is at least 4 since otherwise there would be a vertex having a clique neighbor or a 4-cycle having a chord.
 By  Corollary~\ref{cor_imp} (setting $S'=N[v]$ with $k\geq 4$), we know that  the measure $p$ decreases by at least $\delta_4+q_2\delta_2+(4-q_2)\delta_3+3\delta_3^{<-1>}+\delta_2=4\delta_4+(1-q_2)\delta_3+(q_2+1)\delta_2$.

The above analysis gives the following lemma.

\begin{lemma}
    Step~\ref{S_degree4} followed by applications of reduction rules creates a branching
    vector covered by one of
    $$[\delta_4+4\delta_3, \delta_4+4\delta_3]=[5.624,5.624]~~(q_2=4),$$
     $$[\delta_4+3\delta_3+\delta_3^{<-1>}, 4\delta_4-2\delta_3+4\delta_2]=[5.248,6]~~(q_2=3),$$
    $$[\delta_4+2\delta_3+2\delta_3^{<-1>}, 4\delta_4-\delta_3+3\delta_2]=[4.872,6.624]~~(q_2=2),$$
    $$[\delta_4+1\delta_3+3\delta_3^{<-1>}, 4\delta_4+2\delta_2]=[4.496,7.248]~~(q_2=1),~~\mbox{and}$$
    $$[\delta_4+4\delta_3^{<-1>}, 4\delta_4+\delta_3+\delta_2]=[4.12,7.872]~~(q_2=0).$$
\end{lemma}

Next, we assume the maximum degree of the graph is 3.

\begin{step}[Branching on Other 4-Cycles]\label{b4cycle}
        If there is a 4-cycle $C=v_1v_2v_3v_4$, then branch on the 4-cycle with Branching Rule~\ref{br_4cycle} by excluding either $\{v_1,v_3\}$ or $\{v_2,v_4\}$ from the independent set.
\end{step}

\begin{lemma}
    Step~\ref{b4cycle} followed by applications of reduction rules creates a branching
    vector covered by
    $$
    [6\delta_3-2\delta_2, 6\delta_3-2\delta_2]=[5.248,5.248].
    $$
\end{lemma}

\begin{proof}
All the four vertices in the cycle $C$ are degree-3 vertices or degree-2 vertices since there are no vertices of degree $\geq 4$ now. We can see that there is at most one degree-2 vertex in the cycle $C$.
If there are two nonadjacent degree-2 vertices in a 4-cycle, then there is a twin and R\ref{fold_twin} should be applied. If there are two adjacent degree-2 vertices in a 4-cycle, then either R\ref{fold_d_2_2} or R\ref{4-cycle} can be applied. So there is at most one degree-2 vertex in the cycle $C$.

\textbf{Case 1}: one vertex in $C$, say $v_1$ is of degree 2 and all other vertices in $C$ are of degree 3.

In each branching, all the four vertices in the cycle $C$ will be deleted after applying R\ref{remove_ucf} to R\ref{heavy_vertices}.
    By Corollary~\ref{cor_imp} (setting $S'=\{v_1,v_2,v_3,v_4\}$ with $k=3$), we know that  the measure $p$ decreases by at least $\sum_{u\in\{v_1,v_2,v_3,v_4\}}\delta_{d_u} + 3\delta_3^{<-1>}=3\delta_3 +\delta_2+ 3\delta_3^{<-1>}=6\delta_3-2\delta_2$. We get a branching vector
    $$
    [6\delta_3-2\delta_2, 6\delta_3-2\delta_2].
    $$

\textbf{Case 2}: all the four vertices in $C$ are degree-3 vertices.
 In each branching, all the four vertices in the cycle $C$ will be deleted after applying R\ref{remove_ucf} to R\ref{heavy_vertices}.
    By Corollary~\ref{cor_imp} (setting $S'=\{v_1,v_2,v_3,v_4\}$ with $k=4$), we know that  the measure $p$ decreases by at least $\sum_{u\in\{v_1,v_2,v_3,v_4\}}\delta_{d_u} + 4\delta_3^{<-1>}=4\delta_3 + 4\delta_3^{<-1>}=8\delta_3-4\delta_2$. Note that $8\delta_3-4\delta_2> 6\delta_3-2\delta_2$. So it is covered by the above case.

\end{proof}

From now on, we assume that the graph has a maximum degree 3 and there is no 4-cycle. Next, we will first consider triangles in the graph. For a triangle $C$ in the graph after Step~\ref{b4cycle}, each vertex in $C$ is of degree 3 by Lemma~\ref{lem_no_triangle2} and
is chain-adjacent to a degree-3 vertex not in $C$ by Lemma~\ref{triangle_cycle}.

\begin{step}[Branching on Triangles]
    \label{branchtriangle}
    If there is a triangle $C=v_1v_2v_3$, where we assume without loss of generality that $w(v_1)\geq \max \{w(v_2), w(v_3)\}$ and
    $v_1$ is chain-adjacent to a degree-3 vertex $u\neq v_2, v_3$, then branch on $u$ with Branching Rule 1.
\end{step}

\begin{lemma}
    Step~\ref{branchtriangle} followed by applications of reduction rules creates a branching
    vector covered by one of
$$
    [6\delta_3 -3\delta_2, 7\delta_3+\delta_2]=[4.872,7.376]~~~\mbox{and}
    $$
   $$
    [6\delta_3 -2\delta_2, 5\delta_3+2\delta_2]=[5.248,5.752].
    $$

\end{lemma}

\begin{proof}
We analyze the branching vector by considering the length of the chain between $v_1$ and $u$.

\textbf{Case 1:} $v_1$ and $u$ are adjacent.
In the first branching of excluding $u$ from the independent set, all vertices $C$ will be removed by applying R\ref{remove_ucf} to R\ref{heavy_vertices} since
$w(v_1)\geq \max \{w(v_2), w(v_3)\}$ and $v_2$ and $v_3$ will become unconfined vertices.
By Corollary~\ref{cor_imp} (setting $S'=\{v_1,v_2,v_3,u\}$ with $k=4$), we know that  the measure $p$ decreases by at least $4\delta_3 +4\delta_3^{<-1>}$.

In the second branching of including $u\in S_u$ in the independent set, we will delete $N[u]$ at least.
By Corollary~\ref{cor_imp} (setting $S'=N[u]$ with $k=6-q_2$, where $q_2$ is the number of degree-2 neighbors of $u$), we know that the measure $p$ decreases by at least $6\delta_3-2\delta_2$. We get a branching vector
    $$
    [8\delta_3 -4\delta_2, 6\delta_3-2\delta_2].
    $$

\textbf{Case 2:} the chain between $v_1$ and $u$ is of length 2. We let $w$ be the degree-2 vertex in the chain.
In the first branching of excluding $u$ from the independent set, $w$ will become a degree-1 vertex and then $w$ and $v_1$
will be deleted by applying R\ref{remove_ucf} to R\ref{heavy_vertices}.
By Corollary~\ref{cor_imp} (setting $S'=\{v_1,w,u\}$ with $k=4$), we know that  the measure $p$ decreases by at least $2\delta_3 +\delta_2+4\delta_3^{<-1>}=6\delta_3-3\delta_2$.

In the second branching of including $u\in S_u$ in the independent set, we will delete $N[u]$. Furthermore, $v_2$ and $v_3$ will become unconfined vertices after deleting $N[u]$ and all the three vertices in $C$ will be deleted by applying R\ref{remove_ucf} to R\ref{heavy_vertices}.
Since there is no 4-cycle now, we know that $\{v_1,v_2,v_3\}\cap N[u]=\emptyset$.
By Corollary~\ref{cor_imp} (setting $S'=\{v_1,v_2,v_3\}\cup N[u]$ with $k\geq 4$), we know that  the measure $p$ decreases by at least $4\delta_3 +3\delta_2+ (3\delta_3^{<-1>}+\delta_2)=7\delta_3+\delta_2$. We get a branching vector
    $$
    [6\delta_3 -3\delta_2, 7\delta_3+\delta_2].
    $$

\textbf{Case 3:} the chain between $v_1$ and $u$ is of length 3. We let $w_1$ and $w_2$ be the two degree-2 vertices in the chain.
In the first branching of excluding $u$ from the independent set, $w_1$ and $w_2$ will be removed by reducing degree-1 vertices, and $v_1$ will also be removed since it will become a degree-2 vertex in a triangle.
By Corollary~\ref{cor_imp} (setting $S'=\{v_1,w_1,w_2,u\}$ with $k=4$), we know that  the measure $p$ decreases by at least $2\delta_3 +2\delta_2+4\delta_3^{<-1>}=6\delta_3-2\delta_2$.

In the second branching of including $u\in S_u$ in the independent set, we will delete $N[u]$. Furthermore, $w_1$ and $v_1$ will
also be removed.
By Corollary~\ref{cor_imp} (setting $S'=\{v_1,w_1\}\cup N[u]$ with $k\geq 4$), we know that  the measure $p$ decreases by at least $2\delta_3 +4\delta_2+ (3\delta_3^{<-1>}+\delta_2)=5\delta_3+2\delta_2$. We get a branching vector
    $$
    [6\delta_3 -2\delta_2, 5\delta_3+2\delta_2].
    $$

Note that in a reduced graph, the length of each chain is at most 3. So the above three cases cover all cases.
Since $5\delta_3+2\delta_2< 8\delta_3 -4\delta_2$, we know that $[8\delta_3 -4\delta_2, 6\delta_3-2\delta_2]$ is covered
by $[6\delta_3 -2\delta_2, 5\delta_3+2\delta_2]$.

\end{proof}

Next, we assume that the maximum degree of the graph is at most 3 and all cycles in the graph have a length of at least 5. We still need to deal with some long cycles that contain exactly three degree-3 vertices.

\begin{step}[Branching on Cycles Containing Three Degree-3 Vertices]
    \label{step_cycle_C3}
    If there is a cycle $C$ containing exactly three degree-3 vertices $\{v_1,v_2,v_3\}$, where we assume without loss of generality that $v_1$ is chain-adjacent to a degree-3 vertex $u\neq v_2,v_3$, then branch on $u$ with Branching Rule 1.
\end{step}

\begin{lemma}
    Step~\ref{step_cycle_C3} followed  by  applications  of  reduction  rules  can  create  a branching vector covered by one of
    $$[6\delta_3-4\delta_2,8\delta_3-2\delta_2]=[4.496,7.248],$$
    $$[6\delta_3-3\delta_2,6\delta_3-\delta_2]=[4.872,5.624],~~\mbox{and}$$
    $$[6\delta_3-2\delta_2,6\delta_3-2\delta_2]=[5.248,5.248].$$

\end{lemma}

\begin{proof}
Let $q_2$ denote the number of degree-2 vertices in $N(u)$.  We first consider the branching of
excluding $u$ from the independent set, where we delete $u$ from the graph.
Let $S=\{u\}$.
Recall that $R_S$ is the set of deleted vertices during applying R\ref{remove_ucf}-R\ref{heavy_vertices} on $G-S$. Let $G_{-S}$ be the graph obtained from $G$ by removing $S\cup R_S$ from $G$. We distinguish two cases by considering whether $v_1\in R_S$ or not.

For the case that $v_1\not\in R_S$, we will have that $v_1\in N(S\cup R_S)$ and $v_1$ is left as a degree-2 vertex in $G_{-S}$ since $v_1$ and $u$ are chain-adjacent. Furthermore, we can see that $S\cup R_S$ does not contain any vertex in the cycle $C$,
otherwise all the vertices in $C$ (including $v_1$) should be included in $R_S$ by applying R\ref{remove_ucf}-R\ref{heavy_vertices}. So the cycle $C$ is left in $G_{-S}$.
First, by Lemma~\ref{branch_1 vertex}, we have that $p(G)-p(G_{-S})\geq \delta_3+q_2\delta_2+3\delta_3^{<-1>}=4\delta_3+(q_2-3)\delta_2$.
Second, we consider the cycle $C$ in the remaining graph $G_{-S}$. Now the cycle $C$ contains at most two degree-3 vertices ($v_1$ becomes a degree-2 vertex). If it contains exactly two degree-3 vertices, then
by Lemma~\ref{lem_5cycle_c}, we can further reduce the measure by at least $2\delta_3-\delta_2$ by further applying reduction rules on $G_{-S}$. If $C$ contains at most one degree-3 vertex, then all the vertices in the cycle will be reduced by further applying reduction rules. Thus, we can  further reduce the measure by at least $\delta_3+4\delta_2> 2\delta_3-\delta_2$. In total, the measure $p$ will decrease by at least
$$6\delta_3+(q_2-4)\delta_2.$$

For the case that $v_1\in R_S$,
we apply Corollary~\ref{cor_imp} by letting $S'$ be the set containing $u$, $v_1$, all degree-2 vertices in $N(u)$ and all vertices in the chain between $u$ and $v_1$, and $k=4$.
Now $S'$ contains exactly 2 degree-3 vertices and $q_2$ degree-2 vertices.
We know that the measure $p$ decreases by at least
$$2\delta_3+q_2\delta_2+4\delta_3^{<-1>}=6\delta_3+(q_2-4)\delta_2.$$

So, in the first branching, we can always decrease the measure $p$ by at least
$6\delta_3+(q_2-4)\delta_2$.

Next, we analyze the second branching of including $S_u$ in the independent set, where we will delete $N[u]$ at least. We let $S=N[S_u]$.
We distinguish two cases by considering whether $v_1\in S\cup R_S$ or not.

First, we consider the case that $v_1\notin S\cup R_S$.
Since $v_1\notin S\cup R_S$ and $S=N[S_u]$, we know that $v_1$ is not adjacent to $u$, where $q_2>0$.
As in the analysis for the first branching, we know that all vertices in the cycle $C$ are left in the graph $G_{-S}$, where $v_1$ will become a degree-2 vertex in $G_{-S}$.
By Lemma~\ref{branch_S2}, we know that $p(G)-p(G_{-S})\geq (4-q_2)\delta_3+q_2\delta_2+3\delta_3^{<-1>}=4\delta_3$.
By further reducing the cycle $C$ in $G_{-S}$, the measure will further decrease by at least
$2\delta_3-\delta_2$. In total, the measure will decrease by at least
$$6\delta_3-\delta_2.$$

Second, we consider the case that $v_1\in S\cup R_S$.
Let $S'$ be the set of vertices in $N[u]$ and vertices in the chain between $u$ and $v_1$.

If $u$ and $v_1$ are not adjacent, then $S'$ contains $5-q_2$ degree-3 vertices and at least $q_2$ degree-2 vertices,
 and there are $k=7-q_2$ edges between $S'$ and $V\setminus S'$, where $1\leq q_2 \leq 3$.
 By applying Corollary~\ref{cor_imp}  with $S'$ and $k$, we know that
the measure will decrease by at least
$$(5-q_2)\delta_3+q_2 \delta_2+ 4 \delta_3^{<-1>}\geq 6\delta_3- \delta_2.$$

If $u$ and $v_1$ are adjacent, then $S'$ contains $4-q_2$ degree-3 vertices and $q_2$ degree-2 vertices,
 and there are $k=6-q_2$ edges between $S'$ and $V\setminus S'$, where $0\leq q_2 \leq 2$.
 Note that $S'=N[u]$ now.
For this case, by applying Lemma~\ref{branch_S2} with $S'$ and $k$, we know that
the measure will decrease by at least
$$(4-q_2)\delta_3+q_2 \delta_2+ f(q_2),$$

where $f(q_2)= 4\delta_3^{<-1>}$ if $q_2=2$,  $f(q_2)= 3\delta_3^{<-1>}+2\delta_2$ if $q_2=1$, and
$f(q_2)=3\delta_3^{<-1>}+\delta_2+\delta_3$ if $q_2=0$.

Thus, in this step, we can always branch with one of the following branching vectors
$$[6\delta_3-4\delta_2,8\delta_3-2\delta_2]~~(q_2=0),$$
$$[6\delta_3-3\delta_2,6\delta_3-\delta_2]~~(q_2=1),$$
$$[6\delta_3-2\delta_2,6\delta_3-2\delta_2]~~(q_2=2),~~\mbox{and}$$
$$[6\delta_3-\delta_2,6\delta_3-\delta_2]~~(q_2=3).$$
The last case is covered by the second case. The lemma holds.

\end{proof}

After this step, we can see that the graph has a maximum degree 3 and a minimum degree 2.
The length of any cycle in the graph is at least 5 and each cycle contains at least four degree-3 vertices.
Next, we are going to eliminate degree-3 vertices in the graph according to the following order: first deal with
degree-3 vertices with exactly two degree-2 neighbors (Step~\ref{step_3.1}); then deal with the connected components containing both degree-3 vertices with three degree-2 neighbors and degree-3 vertices with at most one degree-2 neighbors (Step~\ref{step_3.2}); last, all degree-3 vertices in each connected component are either having three degree-2 neighbors or having at most
one degree-2 neighbor, and we deal with these two kinds of connected components separately (Step~\ref{step_3.3} and Step~\ref{step_3.5}).

\begin{step}[Branching on Degree-3 Vertices with Two Degree-2 Neighbors]\label{step_3.1}
    If there is degree-3 vertex $u$ having two degree-2 neighbors and one degree-3 neighbor $v$,
    then branch on $v$ with Branching Rule 1.
\end{step}

\begin{lemma}
    \label{lem_step_9}
Step~\ref{step_3.1} followed by applications of reduction rules creates a branching
    vector covered by one of
    $$[4\delta_3-\delta_2, 8\delta_3-\delta_2]=[3.624,7.624],$$
        $$[4\delta_3, 8\delta_3-4\delta_2]=[4,6.496],~~\mbox{and}$$
    $$[4\delta_3+\delta_2, 6\delta_3]=[4.376,6].$$

\end{lemma}

\begin{proof}
We use $q_2$ to denote the number of degree-2 vertices in $N(v)$, where $0\leq q_2\leq 2$.

In the first branching of excluding $v$ from the independent set, we let $S=\{v\}$.
   Recall that $R_S$ is the set of deleted vertices during applying R\ref{remove_ucf} to R\ref{heavy_vertices} on $G-S$.
   We distinguish two cases by considering whether $u$ is in $R_S$ or not.
   If $u\in R_{S}$, we apply Corollary~\ref{cor_imp} with $S'$ being the set including $u$ and $v$ and all degree-2 neighbors of $u$ and $v$ (note that after deleting $u$ and $v$ all the degree-2 neighbors of them will become degree-1 vertices and will be deleted by applying R\ref{remove_ucf} to R\ref{heavy_vertices}).
   We know that the measure $p$ decreases by at least $$2\delta_3+(q_2+2)\delta_2+4\delta_3^{<-1>}=6\delta_3+(q_2-2)\delta_2.$$
   Otherwise, $u$ is not in $R_S$ and then $u\in N(S\cup R_S)$. Let $G_{-S}$ be the graph obtained from $G$ by removing $S\cup R_S$ from $G$. For the worst case that $R_S=\emptyset$, by  Lemma~\ref{branch_1 vertex},
   we know that $p(G)-p(G_{-S})\geq \delta_3+q_2\delta_2+3\delta_3^{<-1>}$.
   Since $u$ is left in $G_{-S}$, we know that the two degree-2 neighbors of $u$ in $G$
   are also left in $G_{-S}$. Thus, there is a chain containing at least three degree-2 vertices (including $u$) in $G_{-S}$,
   by Lemma~\ref{lem_3d2}, we can further decrease the measure $p$ by at least $2\delta_2$ by applying reduction rules on $G_{-S}$.  In total, the measure $p$ decreases by at least $$\delta_3+q_2\delta_2+3\delta_3^{<-1>}+2\delta_2=4\delta_3+(q_2-1)\delta_2.$$
   Note that $6\delta_3+(q_2-2)\delta_2\geq 4\delta_3+(q_2-1)\delta_2$. In this branching, we can always reduce the measure $p$ by at least $4\delta_3+(q_2-1)\delta_2$.

In the second branching, $v$ is included in the independent set and at least $N[v]$ is deleted.
First, we consider the case that $q_2=0$. We apply Lemma~\ref{branch_S3} with $S'=N[v]$ and $k=6$.
The measure $p$ decreases by at least $4\delta_3+3\delta_3^{<-1>}+\delta_3+2\delta_2=8\delta_3-\delta_2$.
For the case that $q_2=1$, we also apply Lemma~\ref{branch_S3} with $S'=N[v]$ and $k=5$.
The measure $p$ decreases by at least $3\delta_3+\delta_2+5\delta_3^{<-1>}=8\delta_3-4\delta_2$.
For the case that $q_2=2$, we apply Corollary~\ref{cor_imp} with $S'=N[v]\cup N[u]$ and $k=4$.
Now $S'$ contains two degree-3 vertices and
  four degree-2 vertices. The measure $p$ decreases by at least $2\delta_3+4\delta_2+4\delta_3^{<-1>}=6\delta_3$.

Therefore, we can get the three claimed branching vectors.

\end{proof}

\begin{step}[Branching on Degree-3 Vertices of a Mixed Case]
    \label{step_3.2}
    If a degree-3 vertex $u$ without degree-3 neighbors is chain-adjacent to a degree-3 vertex $v$
    with exactly two degree-3 neighbors, then branch on $v$ with Branching Rule 1.
\end{step}

\begin{lemma}
Step~\ref{step_3.2} followed by applications of reduction rules creates a branching
    vector covered by
    $$[4\delta_3, 8\delta_3-2\delta_2]=[4,7.248].$$
\end{lemma}
\begin{proof}
In the first branching of excluding $v$ from the independent set, we let $S=\{v\}$.
 Recall that $R_S$ is the set of deleted vertices during applying R\ref{remove_ucf} to R\ref{heavy_vertices} on $G-S$.
   We distinguish two cases by considering whether $u$ is in $R_S$ or not.
   If $u\in R_{S}$, we apply Lemma~\ref{branch_S3} by letting $S'$ be the set of vertices $N[u]$ and all vertices in the chain from $u$ to $v$ (including $v$) and $k=4$. Now $S'$ contains at most two degree-3 vertices and satisfies the conditions in Lemma~\ref{branch_S3}. We
    know that the measure $p$ decreases by at least
    $$2\delta_3+3\delta_2+4\delta_3^{<-1>}=6\delta_3-\delta_2.$$
   Otherwise, $u$ is not in $R_S$ and then $u\in N(S\cup R_S)$. Let $G_{-S}$ be the graph obtained from $G$ by removing $S\cup R_S$ from $G$. For the worst case that $R_S=\emptyset$, by  Lemma~\ref{branch_1 vertex},
   we know that $p(G)-p(G_{-S})\geq \delta_3+2\delta_3^{<-1>}+ \delta_3=4\delta_3-2\delta_2$.
   Since $u$ is left in $G_{-S}$, we know that the two degree-2 neighbors of $u$ in $G$
   are also left in $G_{-S}$. Thus, there is a chain containing at least three degree-2 vertices (including $u$) in $G_{-S}$,
   by Lemma~\ref{lem_3d2}, we can further decrease the measure $p$ by at least $2\delta_2$ by applying reduction rules on $G_{-S}$.  In total, the measure $p$ decreases by at least $$4\delta_3-2\delta_2+2\delta_2=4\delta_3.$$
   Note that $6\delta_3-\delta_2\geq 4\delta_3$. In this branching, we can always reduce the measure $p$ by at least $4\delta_3$.

In the second branching, $v$ is included in the independent set and at least $N[v]$ is deleted.
We let $S=N[v]$ and consider whether $u\in S\cup R_S$ or not.
If $u\in S\cup R_S$, we apply Corollary~\ref{cor_imp} with $S'$ being the vertex set $N[v]\cup N[u]$ plus all the vertices
in the chain between $u$ and $v$. Now $S'$ contains at least four degree-3 vertices and three degree-2 vertices and $k=6$.  The measure $p$ decreases by at least
$$ 4\delta_3+3\delta_2+(3\delta_3^{<-1>}+\delta_2)=7\delta_3+\delta_2.$$
If $u\not\in S\cup R_S$, then $u\in N(S\cup R_S)$.
We apply Lemma~\ref{branch_S3} by letting $S'=N[v]$ and $k=5$.
Then $p(G)-p(G_{-S})\geq 3\delta_3+ \delta_2+ 5\delta_3^{<-1>}=8\delta_3-4\delta_2$.
Furthermore, $u$ is left in a chain of length at least four in $G_{-S}$.
by Lemma~\ref{lem_3d2}, we can further decrease the measure $p$ by at least $2\delta_2$ by applying reduction rules on $G_{-S}$.  In total, the measure $p$ decreases by at least $$8\delta_3-4\delta_2+2\delta_2=8\delta_3-2\delta_2.$$
Note that $7\delta_3+\delta_2\geq 8\delta_3-2\delta_2$. In this branching, we can always reduce the measure $p$ by at least $8\delta_3-2\delta_2$. We get the claimed branching vector.

\end{proof}

\begin{lemma}\label{twocompnent}
    Let $G$ be the graph after Step~\ref{step_3.2}. For any connected component $H$ of $G$,
    all degree-3 vertices in $H$ either have no degree-3 neighbors or have at least two degree-3 neighbors.
\end{lemma}
\begin{proof}
First, the graph $G$ has no degree-3 vertex with exactly one degree-3 neighbor since Step~\ref{step_3.1} could not be applied now. If there is a degree-3 vertex $u$ having no degree-3 neighbor and a degree-3 vertex $v$
having at least two degree-3 neighbors in a connected component $H$, then there is a path between $u$ and $v$.
We can always choose $u$ and $v$ such that the path between $u$ and $v$ does not contain any degree-3 vertices, i.e., the path is a chain. Thus $u$ and $v$ is chain-adjacent, which means the condition of Step~\ref{step_3.2} holds, a contradiction to the fact that Step~\ref{step_3.2} can not be applied now.

\end{proof}

\begin{step} [Branching on Degree-3 Vertices With At Least Two Degree-3 Neighbors]
    \label{step_3.3}
    If there is a connected component $H$ containing a degree-3 vertex with at least two degree-3 neighbors,
    we let $u$ be the vertex of the maximum weight in $H$ and let $v$ be a degree-3 neighbor of $u$, and branch on $v$ with Branching Rule 1.
\end{step}
 Note that the vertex $u$ of the maximum weight in $H$ can not be a degree-2 vertex, otherwise R\ref{fold_d_2_2}
 can be applied on the degree-2 vertex. So $u$ is a degree-3 vertex.
 By Lemma~\ref{twocompnent}, we know that all degree-3 vertices in $H$ must have at least two degree-3 neighbors, and then we can find a degree-3 neighbor $v$ of $u$, where $v$ also has at least two degree-3 neighbors.

\begin{lemma}\label{l_case3.3}
Step~\ref{step_3.3} followed by applications of reduction rules creates a branching
    vector covered by one of
    $$[4\delta_3-\delta_2, 8\delta_3-\delta_2]=[3.624,7.624],~~\mbox{and}$$
        $$[4\delta_3, 8\delta_3-4\delta_2]=[4,6.496].$$

\end{lemma}
\begin{proof}
    Let $q_2$ be the number of degree-2 neighbors of $v$. Then $q_2=0$ or 1.

    In the first branching of excluding $v$ from the independent set, we let $S=\{v\}$.
    We distinguish two cases by considering whether $u$ is in $R_S$ or not.
    If $u\in R_{S}$, we apply Corollary~\ref{cor_imp} by letting $S'=\{u,v\}$ and $k=4$.
    The measure $p$ decreases by at least $2\delta_3+4\delta_3^{<-1>}=6\delta_3-4\delta_2$.
    If $u\not\in R_{S}$, by Lemma~\ref{branch_1 vertex}, the measure $p$ decreases by
    $\delta_3+(3-q_2)\delta_3^{<-1>}+q_2\delta_3=4\delta_3-(3-q_2)\delta_2$. Furthermore, vertex $u$ is left as a degree-2 vertex. Since $u$ has the maximum weight in $H$, we know that R\ref{fold_d_2_2} can be applied on $u$ to further decrease the measure $p$ by $2\delta_2$. Thus, in this branching, the measure $p$ decreases
    by at least $\min \{4\delta_3-(1-q_2)\delta_2, 6\delta_3-4\delta_2\}=4\delta_3-(1-q_2)\delta_2$ for $q_2=0$ or 1.

    In the second branching, $v$ is included in the independent set and at least $N[v]$ is deleted.
    For the case that $q_2=0$, we apply Lemma~\ref{branch_S3} by letting $S'=N[v]$ and $k=6$.
    The measure $p$ decreases by at least $4\delta_3+\delta_3+2\delta_2+3\delta_3^{<-1>}=8\delta_3-\delta_2$.
    For the case that $q_2=1$, we apply Lemma~\ref{branch_S3} by letting $S'=N[v]$ and $k=5$.
    The measure $p$ decreases by at least $3\delta_3+\delta_2+5\delta_3^{<-1>}=8\delta_3-4\delta_2$.
    So, we get the two claimed branching vectors.

\end{proof}

By Lemma~\ref{twocompnent}, we know that after Step~\ref{step_3.5}, no pair of degree-3 vertices are adjacent.
\begin{step}[Branching on Other Degree-3 Vertices]\label{step_3.5}
    Pick up an arbitrary degree-3 vertex $v$ and branch on it with Branching Rule 1.
\end{step}
\begin{lemma}

Step~\ref{step_3.5} followed by applications of reduction rules creates a branching
    vector covered by
    $$[4\delta_3+6\delta_2,4\delta_3+6\delta_2]=[6.256,6.256].$$
\end{lemma}
\begin{proof}
Let $\{u_1,u_2,u_3\}$ be the three degree-3 chain-neighbors of $v$. By Lemma~\ref{chain_neighbor} we know that
the three degree-3 vertices are different.

In the first branch $S=\{v\}$ and in the second branch $S=S_v\supseteq N[v]$.
Recall that we use $R_S$ to denote the set of deleted vertices during applying R\ref{remove_ucf} to R\ref{heavy_vertices} on $G-S$ and let $G_{-S}=G-(S\cup R_S)$.
In each branch, all vertices
in $N[v]$ will be deleted in $G_{-S}$. If at least one vertex in $\{u_1,u_2,u_3\}$, say $u_1$ is deleted in
$G_{-S}$, then we apply Corollary~\ref{cor_imp} by letting $S'$ being the vertex set $N[v]\cup N[u_1]$ together with all degree-2 vertices in the chain between $v$ and $u$. Then $S'$ contains at least two degree-3 vertices
and five degree-2 vertices, and $k=4$.
The measure $p$ decreases by at least $\sum_{u\in S'}\delta_{d(u)}+4\delta_3^{<-1>}=2\delta_3+5\delta_2+4\delta_3^{<-1>}=6\delta_3+\delta_2$.
If all the three vertices in $\{u_1,u_2,u_3\}$ are left in $G_{-S}$, then all of them will become degree-2 vertices in $G_{-S}$. For this case, we apply Corollary~\ref{cor_imp} by letting $S'=N[v]$ and $k=3$.
The measure $p$ decreases by at least $\sum_{u\in S'}\delta_{d(u)}+3\delta_3^{<-1>}=4\delta_3$.
However, each vertex in $\{u_1,u_2,u_3\}$ has two degree-2 neighbors in $G_{-S}$.
In $G_{-S}$, reduction rules on degree-2 vertices can be applied for at least three times to reduce chains of length $\geq 4$ (even when two vertices in $\{u_1,u_2,u_3\}$ are in the same chain and have a common degree-2 neighbor). So the measure $p$ can be further reduced by $6\delta_2$. In total, the measure $p$ will
decrease by at least $4\delta_3+6\delta_2$. Note that $4\delta_3+6\delta_2 < 6\delta_3+\delta_2$. We
get the claimed branching vector.

\end{proof}

It is easy to see that above steps cover all the cases. Among all the branching vectors, the bottleneck ones are $[4\delta_3,8\delta_3-4\delta_2]=[4,6.496]$ in Lemma~\ref{lem_step_9}, $[4\delta_3+\delta_2,6\delta_3]=[4.376,6]$ in Lemma~\ref{lem_step_9}, and $[4\delta_3, 8\delta_3-4\delta_2]=[4,6.496]$ in Lemma~\ref{l_case3.3}.
All of them have a branching factor of 1.14427.
So we get that

\begin{theorem} \label{result0}
    \textsc{Maximum Weighted Independent Set} can be solved in $O^*(1.1443^p)$ time and polynomial space.
\end{theorem}

By Lemma~\ref{themeasure} and Theorem~\ref{result0}, we get that
\begin{corollary}
\textsc{Maximum Weighted Independent Set} in graphs with average degree at most $x$ can be solved in $O^*(1.1443^{(0.624x-0.872)n})$ time and polynomial space.
\end{corollary}

Let $x=3$ in Lemma~\ref{themeasure}, we get that $p\leq n$ and the following result.

\begin{theorem} \label{result1}
    \textsc{Maximum Weighted Independent Set} in graphs with the average degree at most three can be solved in $O^*(1.1443^n)$ time and polynomial space.
\end{theorem}

\section{Conclusion}

In this paper, we design an exact algorithm for \textsc{Maximum Weighted Independent Set}. With the help of the measure-and-conquer technique, we analyze
a nontrivial running time bound for the algorithm, which has a good performance on sparse graphs.
For graphs with an average degree at most 3.68, our algorithm is even faster then the previous algorithm for degree-4 graphs.
For graphs with an average degree at most three,
the running time bound is $O^*(1.1443^n)$, also improving previous running time bounds for the problem in cubic graphs using polynomial space.
Although the improvement is incremental, such improvements on classic problems have became harder and harder. Any further improvement may need new observations on the structural properties or new techniques to design and analyze the algorithms. For
unweighted \textsc{Maximum Independent Set} on degree-3 graphs,
the running time bound was improved for several times~\cite{DBLP:journals/algorithmica/ChenKX05,xiao05,DBLP:conf/iwpec/BourgeoisEP08,DBLP:journals/jda/Razgon09,DBLP:conf/walcom/Xiao10,DBLP:journals/algorithmica/BourgeoisEPR12,DBLP:journals/tcs/XiaoN13,DBLP:journals/corr/IssacJ13}.
Each improvement is small, but each improvement reveals new properties and new analysis.
Our algorithm is analyzed by the measure-and-conquer technique. The framework of the analysis may also provide a way to analyze other related problems.

\section*{Acknowledgements} The work is supported by the National Natural Science Foundation
of China, under grant 61972070.

%

%
%
%

\end{document}